\pdfoutput=1
\documentclass[aps,prd,amsmath,floats,floatfix, twocolumn,
superscriptaddress,nofootinbib,showpacs]{revtex4-1}

\usepackage[T1]{fontenc}
\usepackage[utf8]{inputenc}
\usepackage{lmodern}
\usepackage{mathrsfs}
\usepackage{tensor}
\usepackage{verbatim}
\usepackage[caption=false]{subfig}
\usepackage[dvipsnames, usenames, table]{xcolor}
\definecolor{linkcolor}{rgb}{0.0,0.3,0.5}
\usepackage[hypertexnames=false, unicode, colorlinks=true, linkcolor=linkcolor,
citecolor=linkcolor, filecolor=linkcolor,urlcolor=linkcolor,
pdfusetitle]{hyperref}
\usepackage{tikz}
\usetikzlibrary{shapes.misc, positioning}

\usepackage[all]{hypcap}
\usepackage{graphicx}
\usepackage{multirow}

\usepackage{xspace}
\usepackage{amssymb}
\usepackage[normalem]{ulem} 
\usepackage{bm} 

\usepackage{microtype}

\graphicspath{%
  {figs/}%
}

\DeclareMathAlphabet{\mathpzc}{OT1}{pzc}{m}{it}
\newcommand{\mycomment}[1]{}
\newcommand{\apjl}{Astrophys. J. Lett.}

\newcommand{\etal}{\textit{et al.\ }}

\begin{document}

\title{Numerical simulations of black hole--neutron star mergers in scalar-tensor gravity}

\newcommand{\Cornell}{\affiliation{Cornell Center for Astrophysics
    and Planetary Science, Cornell University, Ithaca, New York 14853, USA}}
\newcommand{\CornellPhys}{\affiliation{Department of Physics, Cornell
    University, Ithaca, New York 14853, USA}}
\newcommand{\Caltech}{\affiliation{TAPIR 350-17, California Institute of
    Technology, 1200 E California Boulevard, Pasadena, CA 91125, USA}}
\newcommand{\AEI}{\affiliation{Max Planck Institute for Gravitational Physics
(Albert Einstein Institute), D-14476 Potsdam, Germany}}
\newcommand{\UMassD}{\affiliation{Department of Mathematics,
    Center for Scientific Computing and Data Science Research,
    University of Massachusetts, Dartmouth, MA 02747, USA}} 
\newcommand{\UMiss}{\affiliation{Department of Physics \& Astronomy,
    The University of Mississippi, University, MS 38677, USA}}
\newcommand{\Bham}{\affiliation{School of Physics and Astronomy and Institute
    for Gravitational Wave Astronomy, University of Birmingham, Birmingham, B15
    2TT, UK}}
\newcommand\MIT{\affiliation{LIGO Laboratory, Massachusetts Institute of
Technology, Cambridge, Massachusetts 02139, USA}}
\newcommand{\MKI}{\affiliation{Department of Physics and Kavli Institute for Astrophysics and Space Research, Massachusetts Institute of Technology, 77 Massachusetts Ave, Cambridge, MA 02139, USA}}
\newcommand{\CCA}{\affiliation{Center for Computational Astrophysics, Flatiron Institute, New York NY 10010, USA}}
\newcommand{\StonyBrook}{\affiliation{Department of Physics and Astronomy, Stony Brook University, Stony Brook NY 11794, USA}}

\author{Sizheng Ma}
\email{sma@caltech.edu}
\Caltech

\author{Vijay Varma}
\email{vijay.varma@aei.mpg.de}
\thanks{Marie Curie Fellow}
\AEI
\CornellPhys
\Cornell
\UMassD

\author{Leo C. Stein}
\UMiss

\author{Francois Foucart}
\affiliation{Department of Physics \& Astronomy, University of New Hampshire, 9 Library Way, Durham NH 03824, USA}

\author{Matthew D. Duez}
\affiliation{Department of Physics \& Astronomy, Washington State University, Pullman, Washington 99164, USA}

\author{Lawrence E. Kidder}
\Cornell

\author{Harald P. Pfeiffer}
\AEI

\author{Mark A. Scheel}
\Caltech

\hypersetup{pdfauthor={Ma et al.}}

\date{\today}

\begin{abstract}
We present a numerical-relativity simulation of a black hole - neutron star
merger in scalar-tensor (ST) gravity  with binary parameters consistent
with the gravitational wave event GW200115. 
In this exploratory simulation, we consider the Damour-Esposito-Farèse extension to Brans-Dicke theory, and maximize the effect of spontaneous scalarization by choosing a soft equation of state and ST theory parameters at the edge of known constraints.
We extrapolate the gravitational waves, including tensor and scalar
(breathing) modes, to future null-infinity.
The numerical waveforms undergo $\sim22$ wave cycles before the merger, and are in good agreement with
predictions from post-Newtonian theory during the inspiral.
We find the ST system evolves faster than its
general-relativity (GR) counterpart due to dipole radiation, merging a full gravitational-wave cycle before the GR counterpart.  This enables easy differentiation between the ST waveforms and GR in the context of parameter estimation.
However, we find that dipole radiation's effect may be partially degenerate with the NS tidal deformability during the late inspiral stage, and a full Bayesian analysis is necessary to fully understand the degeneracies between ST and binary parameters in GR.
\end{abstract}

\maketitle

\section{Introduction}
\label{sec:introduction}
Increasing numbers of gravitational-wave (GW) events~\cite{TheLIGOScientific:2016pea, LIGOScientific:2018mvr,
LIGOScientific:2020ibl, LIGOScientific:2021djp} have allowed us to probe the
extreme gravity environment near the coalescence of a compact binary system,
which opens up a new chapter for tests of general relativity (GR)
\cite{TheLIGOScientific:2016pea, LIGOScientific:2016lio, Will:2014kxa,
Yunes:2013dva, Yunes:2016jcc, Berti:2015itd, LIGOScientific:2018dkp,
Krolak:1995md, Yagi:2009zz, Ma:2019rei, Carson:2019fxr, Sampson:2014qqa,
Scharre:2001hn, Will:2004xi, Berti:2005qd, Berti:2004bd, Yagi:2009zm,
Arun:2012hf, Cardoso:2011xi, Yunes:2011aa, Berti:2012bp, Tuna:2022qqr}.
To robustly test GR, there is a need for accurate GW predictions both in GR and
beyond-GR theories, so that one can use Bayesian model selection to ascertain
which theory better agrees with GW observations.

Scalar-tensor (ST) theory \cite{jordan1955schwerkraft, fierz1956physical,
PhysRev.124.925, 2008AIPC.1083...34B} is the simplest alternative theory of
gravity, where the strength of gravity is modulated by scalar
field(s).
The original
formulation of ST theory was due to Jordan \cite{jordan1955schwerkraft}, Fierz
\cite{fierz1956physical}, Brans and Dicke
\cite{PhysRev.124.925,2008AIPC.1083...34B} (JFBD), and was generalized by
Bergmann \cite{bergmann1968comments} and Wagoner \cite{Wagoner:1970vr} to
capture more general conformal factors, and by Damour and Esposito-Farèse \cite{Damour:1992we} to
multiple scalar fields.
An important feature of ST theory is scalar radiation, an extra energy
dissipation channel in addition to the usual tensor radiation in GR.
The leading scalar radiation is dipolar, and thus more important at
low frequencies than the quadrupolar waves that control a GR
inspiral~\cite{1989ApJ...346..366W, 1975ApJ...196L..59E,
  Damour:1992we,
Damour:1995kt, Mirshekari:2013vb, Bernard:2018hta, Bernard:2018ivi,
1975ApJ...196L..59E, 1977ApJ...214..826W, Brunetti:1998cc, Will:1994fb,
Lang:2013fna, Sennett:2016klh, Lang:2014osa, Bernard:2022noq, Saijo:1996iz,
Ma:2019rei}.
Under this effect, the evolution of some
strong-gravity systems can deviate from the prediction of GR and leave imprints
on observables. For instance, binary-pulsar systems  have been shown to be a
good laboratory~\cite{1975ApJ...196L..59E, 1977ApJ...214..826W,
1989ApJ...346..366W, PhysRevD.45.1840, Damour:1996ke, Damour:1998jk,
1992Natur.355..132T, Psaltis:2005ai, Freire:2012mg, Shao:2017gwu,
Anderson:2019eay, Archibald:2018oxs, Antoniadis:2013pzd, Zhao:2019suc,
Zhao:2022vig,Gerard:2001fm} (see also Refs.~\cite{Stairs:2003eg, Shao:2016ezh,
Kramer:2016kwa, Wex:2014nva, Esposito-Farese:2004pem, Will:2014kxa,
Kramer:2009zza, Damour:2007uf} for reviews) since the celebrated Hulse-Taylor
PSR B1913+16 \cite{1975ApJ...195L..51H}. By measuring the orbital decay rate of
the systems, one can examine and constrain ST theory via the parametrized
post-Keplerian formalism \cite{Damour:1996ke, PhysRevD.45.1840,
Horbatsch:2011nh, Damour:2007uf}.

The strength of the dipole radiation depends on the scalar charge $\alpha_{\rm NS}$
\cite{Damour:1992we, Damour:1993hw, Anderson:2019hio, Zhao:2019suc},  which
characterizes the ability of an object to condense the scalar field. The
scalar charge of a black hole (BH) vanishes as the no-hair theorems have
been shown to apply in ST \cite{1971ApJ...166L..35T, Sotiriou:2011dz,
Hawking:1972qk, Healy:2011ef, Yunes:2011aa}.  For a binary system, the dipole
radiation power is proportional to its charge difference squared
\cite{1989ApJ...346..366W}: $(\alpha_A-\alpha_B)^2$, where $A$ and $B$ refer to
the two objects in the binary system. Consequently, if two
objects possess similar scalar charges, such as in near equal-mass binary
neutron star (BNS) systems where both stars are similarly scalarized, the dipole
radiation is suppressed.  Conversely, the best tests of ST can come from a
mixed system that consists of a neutron star (NS) and a BH, as only one of them
carries scalar charge. 

While ST theory is strongly constrained in some environments, deviations from GR could also be amplified if a NS undergoes {\emph{spontaneous scalarization}}\footnote{See Refs.~\cite{Barausse:2012da,
Shibata:2013pra, Taniguchi:2014fqa, Palenzuela:2013hsa, Sennett:2016rwa,
Sennett:2017lcx} for two related phenomena: induced and dynamical
scalarization.} in certain conditions \cite{Sotani:2012eb,
Ramazanoglu:2016kul, Rosca-Mead:2020ehn, Barausse:2012da, Shibata:2013pra,
Taniguchi:2014fqa, Palenzuela:2013hsa, Sennett:2016rwa, Sennett:2017lcx,Doneva:2022ewd}, as pointed out by Damour and Esposito-Farèse
\cite{Damour:1993hw,Damour:1996ke}.
At
some critical central density, the equilibrium solutions for NSs' structures bifurcate into several
branches, and the GR branch becomes unstable~\cite{Harada:1997mr,Harada:1998ge}.
The most stable solution corresponds to a scalarized NS with a much larger
scalar charge \cite{Damour:1992we,Damour:1993hw,Anderson:2019hio,Zhao:2019suc}.
Therefore, the dipole radiation and consequential deviations from GR are
significantly amplified in such scalarized BHNS systems, which makes them, if they exist, ideal
environments for studying ST theory.

The LIGO-Virgo detectors~\cite{TheLIGOScientific:2014jea, TheVirgo:2014hva} recently made the landmark observations of the first BHNS binaries via GWs, GW200105 and GW200115
\cite{LIGOScientific:2021qlt}. With the upcoming improvement in GW detector sensitivity \cite{Aasi:2013wya}, including future third-generation detectors \cite{Punturo:2010zz,Hild:2010id,Evans:2016mbw,Reitze:2019iox}, we can look for effects of gravitational dipole radiation at ever-increasing precision. Therefore, it
is timely and vital to give a precise prediction of the evolution of the
scalarized BHNS binaries in ST, especially accurate modeling of their dipole
GW waveforms. Although there have been significant post-Newtonian (PN) efforts
dedicated toward constructing waveforms in ST theory\footnote{See also Refs.~\cite{Khalil:2019wyy,Khalil:2022sii} for an effective-field-theory approach.}
\cite{1989ApJ...346..366W, 1975ApJ...196L..59E, Damour:1992we, Damour:1995kt,
Mirshekari:2013vb, Bernard:2018hta, Bernard:2018ivi, 1977ApJ...214..826W,
Brunetti:1998cc, Will:1994fb, Lang:2013fna, Sennett:2016klh, Lang:2014osa,
Sennett:2016klh, Lang:2014osa, Bernard:2022noq, Saijo:1996iz, Ma:2019rei,
Bernard:2019yfz}, PN theory breaks down as one approaches the merger, or for strongly scalarized NSs. To date, numerical relativity (NR)
still remains the only \emph{ab initio} method to investigate ST
theory near the merger \cite{Harada:1996wt, Scheel:1994yn, Scheel:1994yr, Shibata:1994qd,
10.1143/PTP.49.1195, Novak:1997hw, Healy:2011ef, Barausse:2012da,
Shibata:2013pra, Bezares:2021dma}.  For compact binaries, NR has been used to
simulate binary black holes (BBHs) \cite{Healy:2011ef} and BNSs
\cite{Barausse:2012da, Shibata:2013pra, Bezares:2021dma} in ST. A numerical
simulation of a scalarized BHNS system is still missing. In this work, we aim
to fill this gap by performing fully nonlinear NR simulations
of a BHNS merger in ST theory, with a particular focus on how GW emission is
impacted by spontaneous scalarization. Motivated by the LIGO-Virgo observations, we consider a
GW200115-like system~\cite{LIGOScientific:2021qlt}.

This paper is organized as follows. In Sec.~\ref{sec:eom_numerical} we give a
brief introduction to ST theory and our simulation algorithm. Section
\ref{sec:spontaneous_scalarization} concentrates on our numerical setup and
strategy to maximize the effect of spontaneous scalarization. Section
\ref{sec:numerical_results} provides our major simulation results. Next in
Sec.~\ref{sec:parameter_estimation_bias} we investigate distinguishability
between waveforms in GR and ST, with a particular focus on to what extent the ST
waveform can be mimicked by tidal effects predicted by GR. Finally in
Sec.~\ref{sec:conclusion} we provide some concluding remarks.

Throughout this paper we use the geometric units with $c=G_*=1$, where $G_*$ is
the bare gravitational constant in the Jordan frame. We use the total
Jordan-frame mass to normalize all dimensional quantities (e.g., time and
distance). Meanwhile, we use the Latin letters $a,b,c\ldots$ for spacetime indices, and $i,j,k\ldots$ to represent spatial indices.

\begin{figure*}[tb]
\begin{tikzpicture}
\node (einstein) [draw, rounded corners=20pt,
                 text width=0.3\linewidth,    
                 align=flush center, 
                 inner sep=12 pt,
                 fill=orange!50,
                 label=90:{\textit{\large Gravity Sector}},
                 label=-90:{\textit{\large Pseudospectral}},
                 minimum height=150pt]%
                 at (-1,0)
{\textbf{\Large Einstein Frame}\par
 \large 
$$\bar{G}_{ab}=8\pi (\bar{T}^{\psi}_{ab}+\bar{T}_{ab})$$ \par
$$\bar{\Box}\psi=\frac{1}{2}\frac{d\log\phi}{d\psi}\bar{T}$$};
\node (jordan) [draw, rounded corners=20pt,
                     text width=0.3\linewidth,    
                     align=flush center, 
                     inner sep=12 pt,
                     fill=cyan!30,
                     label=90:{\textit{\large Matter Sector}},
                     label=-90:{\textit{\large Finite Difference}},
                     minimum height=150pt]%
                    at (10.3,0)
{\textbf{\Large Jordan Frame}\par
 \large 
$$\nabla_a(\rho_0 u^a)=0$$ \par
$$\nabla_aT^{ab}=0$$};
\begin{scope}[transform canvas={yshift=25pt}]
\draw[ultra thick, -> ,color=orange ,
     shorten >=10pt,shorten <=10pt
     ] (einstein) -- (jordan) node[midway,above] 
     {$$ \mbox{\Large $g_{ab}=\bar{g}_{ab}/\phi$ } $$};
\end{scope}
\begin{scope}[transform canvas={yshift=-25pt}]
\draw[ultra thick, <- ,color=cyan ,
     shorten >=10pt,shorten <=10pt
     ] (einstein) -- (jordan) node[midway,below] 
     {$$ \mbox{\Large $\bar{T}_{ab}=T_{ab}/\phi$ } $$};
\end{scope}
\end{tikzpicture}
\caption{%
The algorithm of our numerical simulations. We use pseudospectral methods to
evolve the Einstein-frame metric and scalar field, while we use shock-capturing
finite difference to simulate the Jordan-frame matter fields. In practice, we
convert the Einstein-frame metric $\bar{g}_{ab}$ to the Jordan-frame one via
$g_{ab} = \bar{g}_{ab}/\phi$, and then send $g_{ab}$ to the finite difference
domain for hydrodynamics simulations.  Similarly, we transfer the
Jordan-frame stress-energy tensor $T_{ab}$ from the finite difference grid to
the pseudospectral grid, convert it to the Einstein-frame stress-energy tensor
through $\bar{T}_{ab} = T_{ab}/\phi$, and then insert $\bar{T}_{ab}$ into the
Einstein equations in Eqs.~\eqref{Einstein_frame_EOM}.
}
\label{fig:algorithm}
\end{figure*}
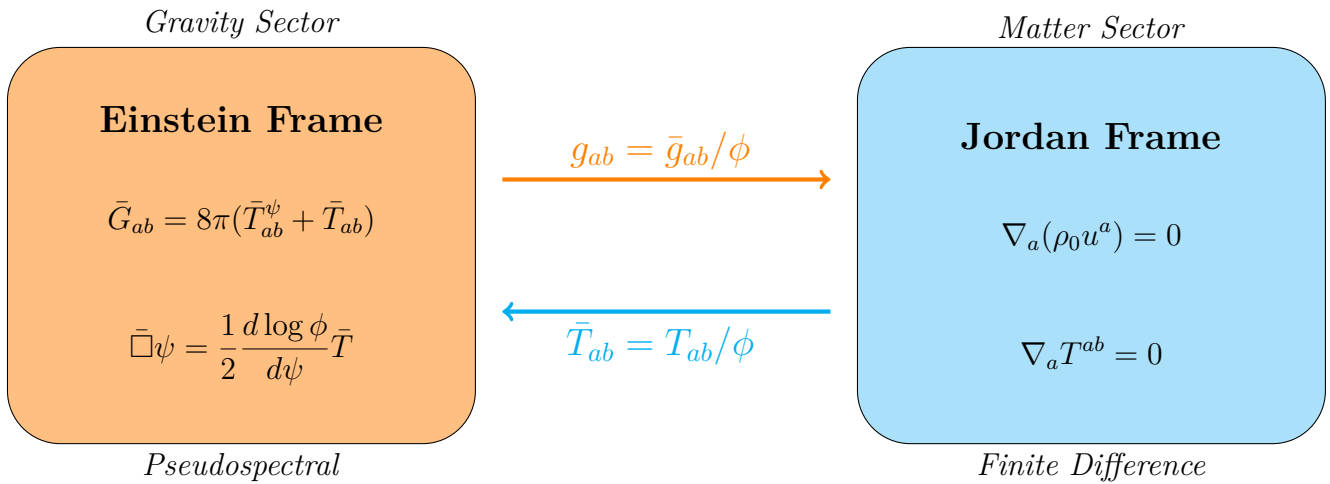

\section{Equations of motion and Numerical methods}
\label{sec:eom_numerical}
In this work we consider a ST theory with a single massless scalar field $\phi$. We first
provide some basic features and equations of motion of this theory in
Sec.~\ref{subsec:jordan-einstein}. Then in
Sec.~\ref{subsec:numerical_algorithm} we introduce our numerical algorithm to
perform the NR simulation. Finally in
Sec.~\ref{subsec:waveform_extrapolation} we provide our method for extrapolating the waveform to future
null infinity.

\subsection{The Jordan and Einstein frames}
\label{subsec:jordan-einstein}
The ST theory is governed by the action \cite{bergmann1968comments,Wagoner:1970vr}
\begin{gather}
    S = \int \! d^4x \frac{\sqrt{-g}}{16 \pi} \! \left [ \phi R
        -\! \frac{\omega(\phi)}{\phi} \nabla_c \phi \, \nabla^c \phi
        \right] \! + \! S_M[g_{ab},\! \Psi_m] ,
\label{eq:jordan_action}
\end{gather}
where $g_{ab}$ is the metric, $g$ is the metric determinant, $R$ is the Ricci scalar, $S_M$ is the action for all matter fields $\Psi_m$, and $\omega(\phi)$ is an arbitrary function of $\phi$ that parameterizes the coupling between the scalar field and metric. The action in
Eq.~\eqref{eq:jordan_action} is written in the \emph{Jordan frame} in
which $\phi$ is nonminimally coupled with the metric $g_{ab}$, whereas
the matter fields are minimally coupled to the metric and not coupled with the scalar field $\phi$, as required by the weak equivalence principle. Therefore, test particles follow the geodesics of the Jordan
frame metric.  NSs are treated as perfect fluids and are governed by the law of conservation of baryon number and energy momentum:
\begin{subequations}
\begin{align}
\nabla_a (\rho_0 \, u^a) = 0, \label{eq:jordan_rho_eq} \\
\nabla_a T^{ab} = 0,
\label{eq:jordan_Tab_eq}
\end{align}
\label{matter_EOM}%
\end{subequations}
where $T^{ab}$ is the stress-energy tensor in the Jordan frame.
The stress-energy tensor for an perfect fluid reads
\begin{align}
  T_{ab} = \rho_0  h u_a  u_b + P g_{ab},
\end{align}
with $\rho_0$ the rest mass density of the fluid, $h$ the specific
enthalpy, $P$ the pressure, and $u_a$ the 4-velocity.

The equations of motion for the metric and the scalar field take complicated
forms in the Jordan frame [see Eq.~(2.6) of Ref.~\cite{Berti:2015itd} for
example]. In particular, the principal symbols of the PDE system is
not diagonal in the $(g_{ab},\phi)$ field space, so it is not
manifestly symmetric-hyperbolic.\footnote{Nevertheless, well-posed formulations of ST theories have been found by Salgado \etal \cite{Salgado:2005hx,Salgado:2008xh}. }  Consequently, the Jordan frame sometimes
is not ideally suited for simulating the metric and scalar
fields. A standard approach to get around this issue is to apply a conformal
transformation \cite{Damour:1992we}: $\bar{g}_{ab} = \phi \, g_{ab}$.  Then the
action becomes:
\begin{gather}
    S = \int \!  d^4x \sqrt{-\bar{g}} \! \left [ \frac{\bar{R}}{16\pi}
        -\! \frac{1}{2} \nabla_c \psi \, \nabla^c \psi
        \right] \! + \! S_M\!\left[\frac{\bar{g}_{ab}}{\phi},\! \Psi_m\right] ,
\label{eq:einstein_action}
\end{gather}
where $\bar{R}$ is the Ricci scalar derived from $\bar{g}_{ab}$, and
\begin{align}
  d\psi = \sqrt{\frac{3 +2 \omega}{16\pi}} ~\frac{d\phi}{\phi}. \label{dpsi_dphi}
\end{align}
The integration of Eq.~\eqref{dpsi_dphi} depends on the form of $\omega(\phi)$, and we will explain more details below in Eqs.~\eqref{phi_psi} and \eqref{omega_psi}.
The transformed metric $\bar{g}_{ab}$ defines a new frame, called the \emph{Einstein frame}; and the scalar field $\psi$ is minimally
coupled in the gravitational sector. The corresponding equations of motion become manifestly symmetric-hyperbolic:
\begin{subequations}
\begin{align}
     & \bar{G}_{ab} =
        8\pi \, (\bar{T}^{\psi}_{ab} + \bar{T}_{ab}),
    \label{eq:einstein_Gab_eq} \\
    &\bar{\Box} \psi = \frac{1}{2} \,\frac{d \log{\phi}}{d \psi} \, \bar{T}. \label{eq:einstein_psi_eq}
\end{align}
\label{Einstein_frame_EOM}%
\end{subequations}
Note that the principal part of the gravitational sector is now identical to its GR counterpart. Here $\bar{G}_{ab}$ is the Einstein tensor obtained from $\bar{g}_{ab}$,
$\bar{T}_{ab} = T_{ab}/\phi$ is the matter stress-energy tensor in the Einstein
frame,
$\bar{T} = \bar{g}^{ab} \, \bar{T}_{ab}$ is its trace, and
$\bar{T}^{\psi}_{ab}$ is the stress-energy tensor of the scalar field, given by
\begin{align}
\bar{T}^{\psi}_{ab} = \nabla_a \psi \, \nabla_b \psi - \frac{1}{2} \,
\bar{g}_{ab} \, \nabla_c \psi \, \nabla^c \psi.
\end{align}
On the other hand, a complication of the Einstein frame is that the hydrodynamic equations gain additional source terms:
\begin{subequations}
\label{eq:hydro_eom_Einstein}
    \begin{align}
    \bar{\nabla}_a \bar{T}^{ab}
    &= - \frac{1}{2} \,\frac{d \log{\phi}}{d \psi} \, \bar{T} \, \nabla^b \psi,
    \label{eq:einstein_matter_eq} \\
    \bar{\nabla_a} (\bar{\rho}_0 \, \bar{u}^a) &=
        -\frac{1}{2} \, \frac{d \log{\phi}}{d \psi}
        \, \bar{\rho}_0 \, \bar{u}^a\, \nabla_a \psi
        \,.
    \label{eq:einstein_baryon_eq}
    \end{align}
\end{subequations}
The scalar field  $\psi$ is now directly coupled with the matter
fields. Because of the source terms on the RHS, particles do not
follow geodesics of $\bar{g}_{ab}$.

\subsection{Numerical algorithm}
\label{subsec:numerical_algorithm}
The single-scalar-field ST theory has been solved numerically for BBHs
\cite{Healy:2011ef} and BNSs \cite{Barausse:2012da,Shibata:2013pra}, with the
pure Einstein frame \cite{Healy:2011ef,Barausse:2012da}, and the pure Jordan
frame \cite{Shibata:2013pra}. In our case, we simulate the BHNS system using
the Spectral Einstein Code (SpEC) \cite{SpECwebsite}, developed by the
Simulating eXtreme Spacetimes (SXS) collaboration \cite{SXSWebsite}. SpEC
adopts the generalized harmonic formalism \cite{Lindblom:2005qh}, where the
Einstein equations are cast into a first-order symmetric hyperbolic (FOSH) form.
It is ideal to use SpEC to evolve the metric and the scalar field sectors in
the Einstein frame [Eqs.~\eqref{Einstein_frame_EOM}]. The reason is twofold.
(a) The equations of motion in the Einstein frame are manifestly symmetric-hyperbolic, as mentioned in
Sec.~\ref{subsec:jordan-einstein}. Therefore the well-posedness of the Cauchy
problem is straightforwardly established. (b) The principal parts of
Eqs.~\eqref{Einstein_frame_EOM} are identical to that of GR with a
Klein-Gordon field. Consequently, we can utilize the existing GR FOSH
system~\cite{Lindblom:2005qh} and the FOSH system for scalar
fields~\cite{Holst:2004wt,Scheel:2003vs}
to perform the simulations.

For the hydrodynamics, one could in principle
approach the problem in the same Einstein frame by evolving
Eqs.~\eqref{eq:hydro_eom_Einstein}. But this will complicate the problem
because the extra source terms in Eqs.~\eqref{eq:hydro_eom_Einstein}, which depend
on the scalar field, need to be added to the existing hydrodynamic code
infrastructure in SpEC \cite{Duez:2008rb}. Furthermore, any routine in SpEC
that assumes the simple form of energy-momentum and Baryon number conservation
in Eq.~\eqref{matter_EOM} will need to be
revisited. To save the amount of
code changes required, here we propose a simpler algorithm to fulfill the
goal.

We adopt a hybrid scheme, illustrated in
Fig.~\ref{fig:algorithm}.
We evolve the hydrodynamic system in the
Jordan frame, where the corresponding equations [Eq.~\eqref{matter_EOM}]
are the same as their GR counterparts due to the weak equivalence principle.
This lets us use the entire relativistic hydrodynamics module without
modification. Meanwhile, we use the FOSH systems to treat the metric and the
scalar field in the Einstein frame. Since the Jordan and Einstein frames are
related, a proper data flow needs to be established to evolve them together.
An essential step is to pass the Jordan-frame metric $g_{ab}$ and stress-energy
tensor $T_{ab}$ back and forth (see
App.~\ref{app:two_grid_transformation} for details): The
Einstein-frame metric
$\bar{g}_{ab}$ is converted to its Jordan-frame version $g_{ab}$ via $g_{ab} =
\bar{g}_{ab}/\phi$, then $g_{ab}$ is sent to the Jordan frame for evolving the hydrodynamics. Similarly, the Jordan-frame stress-energy tensor $T_{ab}$ is
converted to the Einstein-frame one through $\bar{T}_{ab} = T_{ab}/\phi$, and
inserted into the Einstein equations in Eqs.~\eqref{Einstein_frame_EOM}.

Within SpEC, this communication is made easier by the two-grid method already
used in hydrodynamics simulations~\cite{Duez:2008rb}, wherein the metric
sector is evolved on a pseudospectral grid, while the hydrodynamic equations
are evolved on a finite difference grid that can handle shocks. At each time
step, the metric from the pseudospectral grid is already interpolated
onto the finite
difference grid and is fed to the hydrodynamic
equations, and the matter fields are passed by interpolation from the finite difference grid to
the pseudospectral grid and are fed to the
stress-energy tensor in the Einstein equations. For the ST simulations, the
metric and the scalar field are evolved in the Einstein frame [see
Eqs.~\eqref{Einstein_frame_EOM}] on the pseudospectral grid, but before the metric is interpolated to the finite difference grid, it is first converted to
the Jordan frame. Similarly, the hydrodynamics equations [see
Eqs.~\eqref{matter_EOM}] are evolved in the Jordan frame, but before the matter terms are transformed to the Einstein frame, they are first interpolated to the pseudospectral grid.

\begin{table*}
    \centering
\renewcommand\arraystretch{1.5}
    \caption{Summary of the parameters of the GW200115-like BHNS system we consider. The NS has a
baryonic mass $m^{\rm B}$ and a Jordan-frame mass $m^{\rm J}_{\rm NS}$. Its
radius in the Jordan frame is $R_{\rm
ST}^{\rm J}$. In the absence of the scalar field, its radius is
$R_{\rm GR}$, and $C_{\rm GR}=m^{\rm J}_{\rm NS}/R_{\rm GR}$ is its
compactness. The GR tidal Love number of the NS is $k_2^{\rm GR}$; $\Lambda_2^{\rm GR}$ is the corresponding tidal
deformability; $\alpha_{\rm NS}$ is its scalar charge.  To maximize the effect
of spontaneous scalarization, we choose $(\beta_0, \alpha_0)=(-4.5,
-3.5\times10^{-3})$. The BH has a Jordan-frame mass $m^{\rm J}_{\rm BH}$.
 Its dimensionless spin along is denoted by $\bm{\chi}^{\rm BH}_{\rm
   init}$ and is anti-aligned with the Newtonian angular momentum
 direction $\hat{\bm{L}}_{N}$. The mass-weighted tidal deformability of the BNHS system is $\tilde{\Lambda}_2^{\rm GR}$. $R_{\rm bdry}$ indicates the radius of the simulation domain, in the unit of total mass $M=7.2M_\odot$, and $N_{\rm cycle}$ is the number of orbital cycles before merger.
The remnant is a BH with mass $m_f$ and
spin $\chi_f$, where $m_f$ is in the unit of $M$.}
\begin{tabular}{>{\centering\arraybackslash}p{2cm}@{\hspace{0cm}}>{\centering\arraybackslash}p{2cm}@{\hspace{0cm}}>{\centering\arraybackslash}p{2cm}@{\hspace{0cm}}>{\centering\arraybackslash}p{2cm}@{\hspace{0cm}}>{\centering\arraybackslash}p{2cm}@{\hspace{0cm}}>{\centering\arraybackslash}p{2cm}@{\hspace{0cm}}>{\centering\arraybackslash}p{2cm}@{\hspace{0cm}}>{\centering\arraybackslash}p{2cm}}
\hline\hline
$m^{\rm B}/M_{\odot}$ & $m^{\rm J}_{\rm NS}/M_{\odot}$ & $\chi^{\rm NS}_{\rm init}$ & $R_{\rm ST}^{\rm J}$/km & $R_{\rm GR}$/km & $C_{\rm GR}$ & $k_2^{\rm GR}$ & $\Lambda_2^{\rm GR}$  \\ 
  \rowcolor{lightgray} 1.71 & 1.5 & 0.0 & 10.58 & 10.55 & 0.21 & 0.0803 & 131.1   \\ \\ \hline \hline
    $\alpha_{\rm NS}$ & $m^{\rm J}_{\rm BH}/M_{\odot}$& $\bm{\chi}^{\rm BH}_{\rm init}$ & $\tilde{\Lambda}_2^{\rm GR}$ & $R_{\rm bdry}/M$ & $N_{\rm cycle}$ & $m_f/M$& $\chi_f$    \\
     \rowcolor{lightgray} 0.18 & 5.7& $-0.19 \hat{\bm{L}}_{N}$ & 2.95 &  500 & 12 & 0.98 & 0.38
\end{tabular}
     \label{table:ns_properties}
\end{table*}

\subsection{Waveform extrapolation}
\label{subsec:waveform_extrapolation}
One of the most important tasks of our numerical simulations is to compute GWs
at future null infinity, where we approximate GW detectors to reside. Methods have
been developed, including wave extrapolation \cite{Boyle:2009vi,Iozzo:2020jcu}
and Cauchy-Characteristic Extraction (CCE) \cite{Moxon:2020gha,Moxon:2021gbv},
to extract the GWs from simulations with finite domains. This paper
adopts the extrapolation method and leaves CCE for future work.

Following the standard treatment
in PN theory
\cite{Mirshekari:2013vb,Lang:2013fna,Sennett:2016klh,Bernard:2022noq}, we
define a new conformally transformed metric $\tilde{g}_{ab}$ by
\begin{align}
    &\tilde{g}_{ab}=(\phi/\phi_0) g_{ab}=\bar{g}_{ab}/\phi_0,
\end{align}
which differs from the Einstein frame metric $\bar{g}_{ab}$ by a factor of
$\phi_0$, the asymptotic value of the scalar field. The factor is
introduced so that the metric $\tilde{g}_{ab}$ takes its Minkowski form
$\eta_{ab}\equiv {\rm diag}(-1,1,1,1)$ far from the system.  In our simulations, we find that the value of $\phi_0$ is always close to 1, and the difference is negligible, so we will not
distinguish $\tilde{g}_{ab}$ from $\bar{g}_{ab}$ below. The gravitational
perturbation $\tilde{h}_{ab}$ associated with $\tilde{g}_{ab}$ is given by
\begin{align}
    \tilde{h}^{ab}=\eta^{ab}-\sqrt{-\tilde{g}} \, \tilde{g}^{ab},
\end{align}
whose indices are raised and lowered by $\eta^{ab}$.
Then the Jordan-frame metric can be written as \cite{Lang:2013fna}
\begin{align}
    g_{ab}=\eta_{ab}+\tilde{h}_{ab}-\frac{1}{2}\tilde{h}\eta_{ab}-\Psi\eta_{ab}+\mathcal{O}\left(\frac{1}{r^2}
    \right), \label{eq:metric_scalar_tensor_decomp}
\end{align}
where 
\begin{align}
    \Psi=\frac{\phi-\phi_0}{\phi_0}. \label{eq:Psi_phi0}
\end{align}
Due to the equation of geodesic deviation
\cite{poisson2014gravity}, the GW measured by a detector corresponds to the
components of the Riemann curvature tensor,
\begin{align}
    R_{0i0j}=-\frac{1}{2}\ddot{\tilde{h}}_{ij}^{\rm TT}-\frac{1}{2}\ddot{\Psi}(\hat{N}_i\hat{N}_j-\delta_{ij}), \label{eq:Riemann_GW}
\end{align}
where ``TT'' refers to the transverse-traceless projection of $\tilde{h}_{ij}$,
and $\hat{N}_i$ is GW's propagation direction.
As a result, the tensor
field $\tilde{h}_{ij}^{\rm TT}$ contributes to the $+$ and $\times$
polarizations of the GW signal as in GR, while the scalar field $\Psi$
corresponds to a transverse breathing mode.\footnote{Longitudinal and
vector polarizations vanish in ST gravity~\cite{poisson2014gravity}.}

To extract the three GW polarizations from our numerical simulations, we notice that the gravitational perturbation $\tilde{h}_{ab}$ is
associated with the Einstein-frame metric $\bar{g}_{ab}$, so we can restrict
ourselves to this frame during the extrapolation.
On the scalar sector side, $\psi$ [defined in
Eq.~\eqref{dpsi_dphi}] is our evolved variable in the Einstein frame. We can convert it to the observable $\Psi$ by integrating Eq.~\eqref{dpsi_dphi} and then inserting the result into Eq.~\eqref{eq:Psi_phi0}. Note that the integration depends on the form of $\omega(\phi)$ and we will provide more details in Eq.~\eqref{phi_psi}. In practice, we first measure the values of $\tilde{h}$ and $\psi$ at multiple extraction radii at each timestep, and then extrapolate their values to null infinity
$\mathscr{I}^+$.
For each radius, we decompose
$\tilde{h}=\tilde{h}_+-i\tilde{h}_\times$ and $\psi$ into a sum over a
set of (spin-weighted) spherical harmonics
${}_{s}Y_{lm}(\iota,\varphi)$,
\begin{subequations}
\begin{align}
&r\tilde{h}/M
=\sum_{l,m}\tensor[_{-2}]{{Y_{lm}}}{}(\iota,\varphi)\tilde{h}_{lm}
+\mathcal{O}(r^{-1}),
\label{h22_extrapolate}\\
&r\psi/M =\sum_{l,m}Y_{lm}(\iota,\varphi)\psi_{lm} +\mathcal{O}(r^{-1}),
\label{psi_extrapolate}
\end{align}
\end{subequations}
where we used the fact that $\tilde{h},\psi\sim 1/r$ in the wave
zone. Each field $\tilde{h}_{lm}$ and $\psi_{lm}$ is extrapolated to $\mathscr{I}^+$ following
the algorithm outlined in Refs.~\cite{Iozzo:2020jcu,Boyle:2013nka,Boyle:2014ioa,Boyle:2015nqa}, with the $\texttt{PYTHON}$ package $\texttt{scri}$ \cite{scri,mike_boyle_2020_4041972}. In particular, the null
rays are parameterized by an approximate retarded time $u$, given by
\begin{align}
    u=t_{\rm corr}-r_*,
\end{align}
with
\begin{align}
    r_*=r+2M^{\rm E}\log\left(\frac{r}{2M^{\rm E}}-1\right),
\end{align}
where $M^{\rm E}=m_{\rm NS}^{\rm E}+m_{\rm BH}^{\rm E}$ is the total Einstein-frame
mass, and we refer to
Refs.~\cite{Boyle:2009vi,Iozzo:2020jcu} for the expression of the corrected
time $t_{\rm corr}$.
Finally these fields are interpolated to common sets of $u$ and
fit in powers of $1/r$, allowing to approximate the $r\to\infty$ limit.

\section{Binary and scalar parameters}
\label{sec:spontaneous_scalarization}
In Sec.~\ref{subsec:binary_parameter}, we provide the binary parameters we consider for the
BHNS system, which are chosen to be consistent with GW200115
\cite{LIGOScientific:2021qlt}. Then in Sec.~\ref{subsec:scalar-tensor}, we
introduce our strategy for choosing the parameters of the scalar field and the
NS. As mentioned in Introduction, a NS can undergo significant scalarization
under certain conditions, leading to nonnegligible dipole radiation while the
scalarized NS orbits in the binary system. This extra energy dissipation
channel accelerates the evolution of the BHNS system and thus makes the emitted
GWs distinguishable from their GR counterparts. In our simulations, we want to
highlight such distinctions by optimally picking the ST
theory parameters and the EOS of the NS.

\subsection{The binary parameters}
\label{subsec:binary_parameter}
We summarize the parameters of the GW200115-like BHNS system
\cite{LIGOScientific:2021qlt} we consider in Table~\ref{table:ns_properties}. The binary
system consists of a nonrotating NS with a Jordan-frame mass $m^{\rm J}_{\rm
NS}$ of $1.5M_\odot$, and a spinning BH with $m^{\rm J}_{\rm BH}=5.7M_\odot$.
The dimensionless spin of the BH $\chi^{\rm BH}_{\rm init}$ is $-0.19$, i.e. it is anti-aligned with the orbital angular momentum. We set the initial separation
between the BH and the NS $D_{\rm init}$ to $11.7M$, where $M\equiv m_{\rm
BH}^{\rm J}+m_{\rm NS}^{\rm J}=7.2M_\odot$ is the total Jordan-frame mass; and place the outer boundary of the system at $R_{\rm bdry}=500M$. The
system undergoes $N_{\rm cycle}\sim 12$ cycles prior to the merger. The orbital
eccentricity is reduced iteratively to $e_{\rm orb}\sim 1.6\times10^{-4}$
\cite{Buonanno:2010yk}.

Due to our two-grid method described in Fig.~\ref{fig:algorithm}, the NS
resides in the Jordan frame while the BH is in the Einstein frame. So in
practice one needs to specify the Einstein-frame mass of the BH $m^{\rm E}_{\rm
BH}$ instead, which is related to the Jordan-frame mass $m^{\rm J}_{\rm BH}$
through
\cite{Damour:1992we}
\begin{align}
    m^{\rm E}_{\rm BH}=\frac{m^{\rm J}_{\rm BH}}{\sqrt{\phi}}, \nonumber
\end{align}
where $\phi$ is evaluated at the position of the BH. We find that
$|\phi-1|\lesssim 5\times 10^{-5}$ in the vicinity of the BH, during the inspiral stage, therefore the
difference between $m_{\rm BH}^{\rm J}$ and $m_{\rm BH}^{\rm E}$ is negligible; thus we simply set $m^{\rm E}_{\rm BH}=5.7M_\odot$.

\subsection{The parameters of the scalar field and the NS}
\label{subsec:scalar-tensor}
For a given Jordan-frame mass $m_{\rm NS}^{\rm J}$, the strength of spontaneous
scalarization for the NS depends on $\omega(\phi)$, as well
as the EOS and compactness \cite{Yagi:2021loe,Anderson:2019hio}. To look for
the optimal choices to maximize the scalarization in our BHNS simulation, we
consider a single Tolman–Oppenheimer–Volkoff (TOV) NS in an isolated gravity environment and investigate the
impact of the scalar field on the stellar internal structure.

The function $\omega(\phi)$ characterizes the coupling between the scalar
field and gravity. In this work we follow
Ref.~\cite{PhysRevLett.70.2220}, whose idea was to Taylor expand the
coupling function $\ln\phi$ in $\psi$,
\begin{gather}
    \phi = \exp{\left[-4\sqrt{\pi}\alpha_0 \, (\psi-\psi_0)-4 \pi \, \beta_0 \, (\psi-\psi_0)^2 \right]}. \label{phi_psi}
\end{gather}
Using Eq.~\eqref{dpsi_dphi}, we obtain
\begin{align}
\omega=\frac{1}{2}\frac{1}{[\alpha_0+\sqrt{4\pi}\beta_0(\psi-\psi_0)]^2}-\frac{3}{2}. \label{omega_psi}
\end{align}
Here $\psi_0$ is the asymptotic value of $\psi$ that can also be associated with cosmological expansion \cite{Anderson:2016aoi,Steinhardt:1994vs,Boisseau:2000pr}. For simplicity, we follow Ref.~\cite{Barausse:2012da} and set $\psi_0=0$. The other two constants
$\alpha_0$ and $\beta_0$ determine the features of the ST theory. In particular,
if $\beta_0=0$ we get the JFBD theory \cite{jordan1955schwerkraft,
fierz1956physical, PhysRev.124.925, 2008AIPC.1083...34B}, which is
parameterized by $\alpha_0=-(3+2\omega_{\rm BD})^{-1/2}$, where $\omega_{\rm
BD}$ is the Brans-Dicke (BD) parameter. In the low-density solar system environment,
its value is severely restricted to $\omega_{\rm
BD}>40000$ by the Cassini mission \cite{Will:2014kxa,2003Natur.425..374B},
which corresponds to $|\alpha_0|\lesssim 3.5\times10^{-3}$. In
addition, current binary pulsar measurements place a constraint
$\beta_0\gtrsim-4.5$, because no spontaneous scalarization has been
detected yet \cite{Berti:2015itd}. See also
Refs.~\cite{Shao:2017gwu,Anderson:2019eay,Archibald:2018oxs, Zhao:2022vig} for more recent
updates.

\begin{figure}[tb]
        \includegraphics[width=\columnwidth,clip=true]{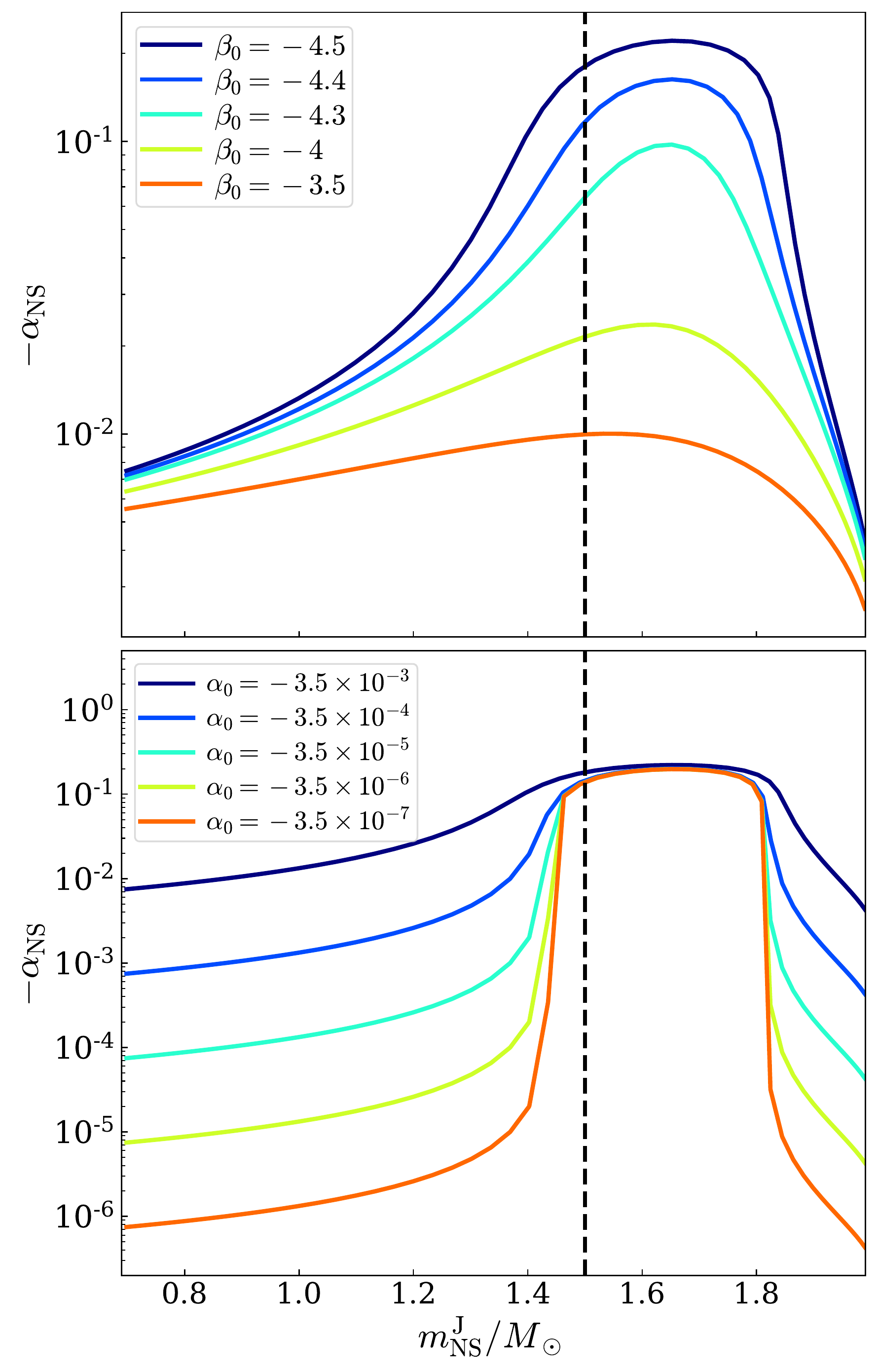}
\caption{The scalar charge of a NS as a function of $m^{\rm J}_{\rm NS}$, with
a variety of $\alpha_0, \beta_0$. Upper panel: varying $\beta_0$ while $\alpha_0 = -3.5\times10^{-3}$; lower panel: varying $\alpha_0$ with $\beta_0 = -4.5$.
The EOS is summarized in Table~\ref{table:ns_properties}, which has been selected to amplify the scalarization. The vertical dashed lines correspond to the NS in our simulation
($m^{\rm J}_{\rm NS}=1.5M_\odot$). We choose $\psi_0=0$ in both
panels.
}
 \label{fig:scalar_charge}
\end{figure}

As pointed out by Damour and Esposito-Farèse
\cite{PhysRevLett.70.2220,Damour:1996ke}, even though a scalar-tensor theory
with $|\alpha_0|\ll 1$ is indistinguishable from GR within the weak-gravity
regime, a negative value of $\beta_0$ can lead to significant relativistic
deviations in a strong-gravity environment, such as spontaneous scalarization
of a NS. The size of the scalarization is characterized by the scalar charge $\alpha_{\rm NS}$ \cite{PhysRevLett.70.2220,Damour:1992we}. In this paper, we adopt the definition of
$\alpha_{\rm NS}$ from Refs.~\cite{PhysRevLett.70.2220,Damour:1992we}, which
differs from the convention used by the PN community by a minus sign (see
App. A of Ref.~\cite{Sennett:2016klh} for translating notation); consequently, we have $\alpha_{\rm NS}<0$.
For a Newtonian star, $\alpha_{\rm NS}$ reduces to $\alpha_0$; thus is
independent of its internal structure (a proof can be found in App.
\ref{app:stellar_structure}). For a strongly self-gravitating scalarized star, its structure is governed by the TOV equation with an extra scalar field,
see e.g.\ Eqs.~$(7-9)$ of Ref.~\cite{PhysRevLett.70.2220}; we provide
a brief review in App. \ref{app:stellar_structure}. We numerically solve the TOV equation, and the choice of the EOS will be discussed shortly. Then we compute the corresponding scalar charge
$\alpha_{\rm NS}$ with Eq.~\eqref{eq:scalar_charge_def}.
Figure \ref{fig:scalar_charge} shows
$\alpha_{\rm NS}$ as a function of the Jordan-frame mass $m^{\rm J}_{\rm
NS}$, using a variety of $\beta_0$ (the upper panel, with $\alpha_0$ being fixed to $-3.5\times10^{-3}$) and $\alpha_0$ (the lower panel, with $\beta_0$ being fixed to $-4.5$) values.
Notice that sharp transitions develops at $m^{\rm J}_{\rm NS}\sim1.4M_\odot$ and
$1.8 M_\odot$ as $\alpha_0\to
0$.  The NSs between these masses are spontaneously scalarized. In addition, we see the scalar charge increases with the absolute value
of $\alpha_0$ and $\beta_0$ for a fixed $m^{\rm J}_{\rm NS}$ (e.g. the vertical dashed line). Therefore, we
chose $(\beta_0, \alpha_0)=(-4.5, -3.5\times10^{-3})$ below to  maximize
the effect of scalarization.

On the other hand, we can also leverage the freedom of choosing an EOS to magnify the scalarization. Here we restrict ourselves to the spectral EOSs provided in \cite{Foucart:2019yzo}, which allows a broad range of cold and beta-equilibrium EOSs (see Fig.~1 of Ref.~\cite{Foucart:2019yzo}). The parametrization reads
\begin{align}
    P(\rho)=
    \begin{cases}
    \kappa_0\rho^{\Gamma_0}, &  \rho<\rho_0, \\
    P_0 \exp\left[\int_0^x\Gamma(\tilde{x})\,d\tilde{x}\right],& \rho>\rho_0,
    \end{cases}
\end{align}
with $\rho_0$ a reference density, $P_0=P(\rho_0)$,
$\Gamma(x)=\gamma_2x^2+\gamma_3x^3$ and $x=\ln(\rho/\rho_0)$. Among the options, we find the following soft EOS that gives rise to the strongest scalarization effect (obtained from Table~III of
Ref.~\cite{Foucart:2019yzo}): 
\begin{align*}
    &\Gamma_0=2,\quad \rho_0=8.44019\times10^{-5},\quad P_0=1.20112\times10^{-7} \\
    &\gamma_2=0.475296,\quad \gamma_3=-0.117048.
\end{align*}
Note that $\rho_0$ and $P_0$ are in $G_* = c = M_\odot = 1$ units. This specific EOS can produce macroscopic properties that are
compatible with current constraints, including the mass-radius relation, tidal
deformability, and maximum NS mass~\cite{Foucart:2019yzo}. However, it should
be noted that this EOS lacks composition and temperature dependence \cite{Foucart:2019yzo}, which
makes it less realistic in those aspects.

For comparison, we also solve a NS with the same Jordan-frame mass in GR, and summarize the corresponding stellar properties in Table~\ref{table:ns_properties}. The compactness of the NS is
$C_{\text{GR}}\sim 0.22$, with a tidal Love number $k_2^{\rm GR}$ of 
$\sim0.08$ \cite{Hinderer:2007mb} and a tidal deformability $\Lambda_2^{\rm GR}=\frac{2}{3}\frac{k_2^{\rm GR}}{C_{\rm GR}^5}$ of
$\sim 131.1$
\cite{Flanagan:2007ix} in the absence of the scalar field.

To end this section, we emphasize that our choices for the EOS and the ST theory parameters are intentionally made to produce a large scalar field: the values of $(\alpha_0,\beta_0)$ lie on the edge of existing constraints \cite{Will:2014kxa,2003Natur.425..374B,Berti:2015itd,Shao:2017gwu,Anderson:2019eay,Archibald:2018oxs}, even though they may not be preferred in the actual astrophysical environment. The current idealized configuration is to justify our simulation code and to investigate the maximum possible detectability of the dipole radiation emitted by BHNS systems. Future work is being planned to explore more moderate scenarios.

\begin{figure}[tb]
        \includegraphics[width=\columnwidth,clip=true]{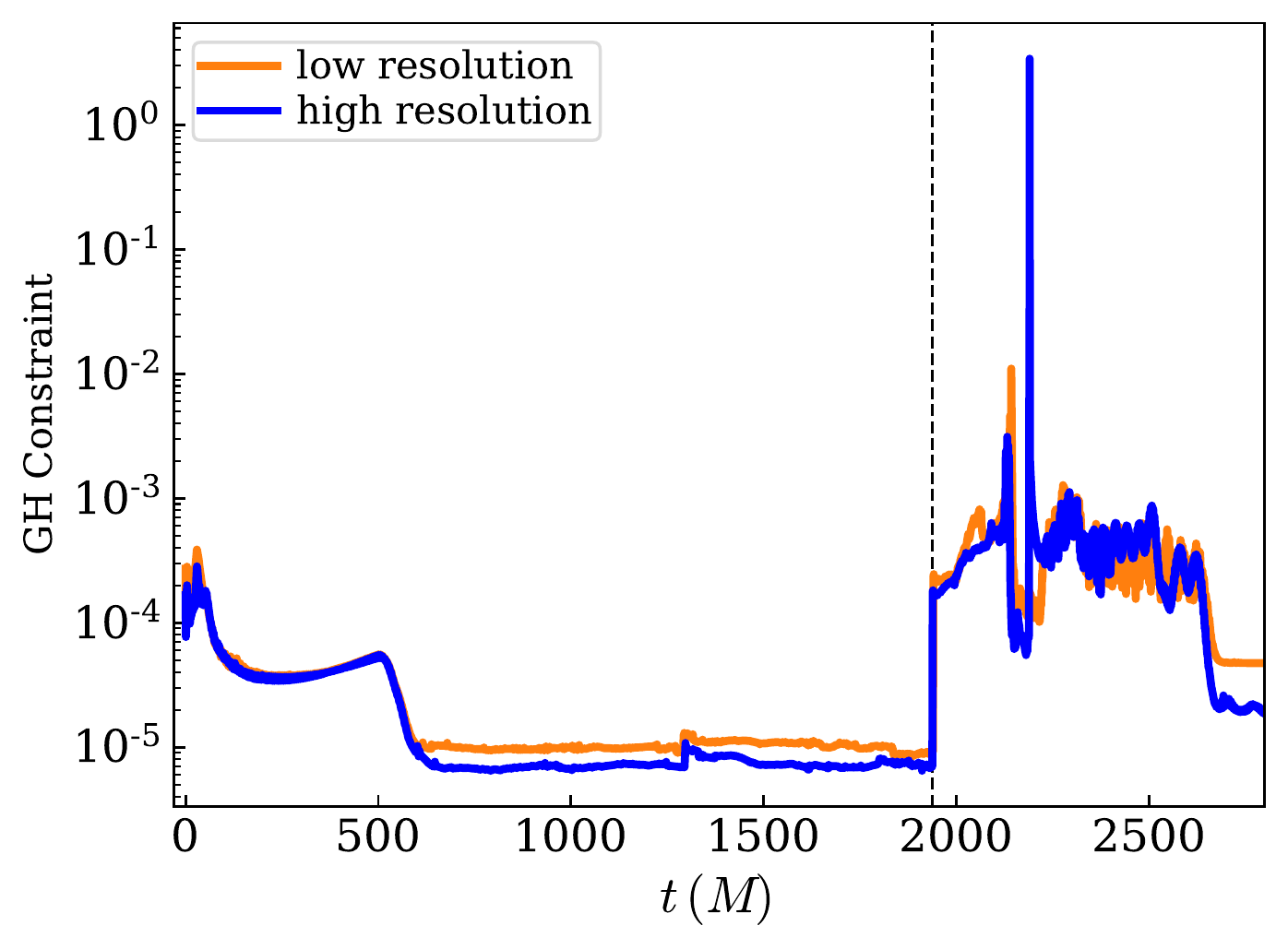}
\caption{
The evolution of the volume-weighted constraint energy for the metric, evolved with GR. The orange (blue) curve
corresponds to the low (high) resolution. The vertical dashed line indicates the onset of the merger.
}
\label{fig:constraint_gr}
\end{figure}

\begin{figure*}[tb]
        \includegraphics[width=\textwidth,clip=true]{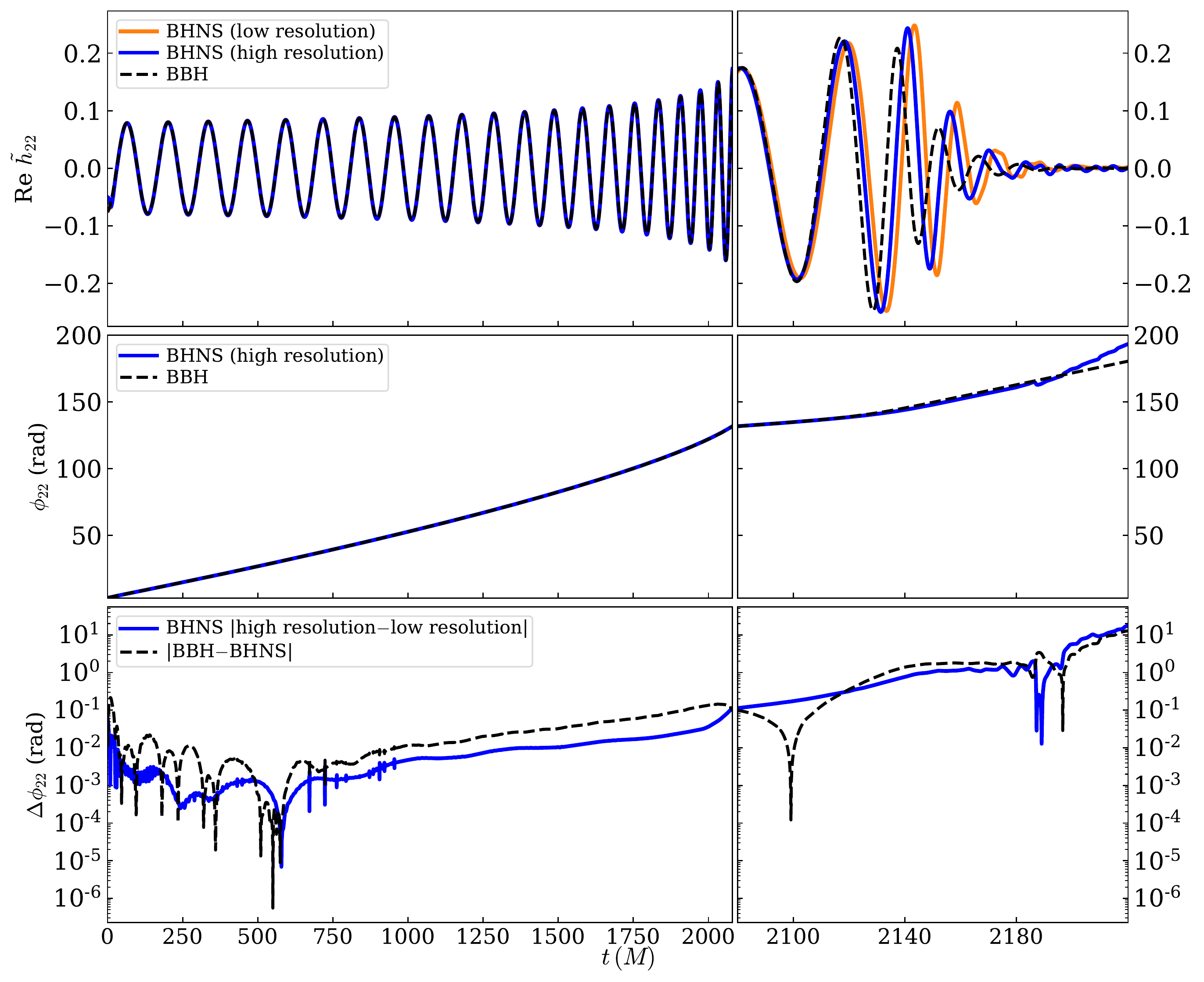}
\caption{Upper panels: The GW harmonic $\tilde{h}_{22}$ of the BHNS system
evolved with GR, using a low (in orange) and high resolution (in blue). Two BHNS
waveforms are compared to that of the BBH system (in black) which has the same
mass ratio and spins. We align the three waveforms by minimizing their mismatch over time and phase shifts, with the optimization window chosen to be $[200M,800M]$. Middle panels: the GW phases of the high-resolution BHNS binary (in blue) and the BBH binary (in black). Lower panels: the GW phase difference between the BBH and
the BHNS system (in black). It is compared to the numerical resolution difference of the BHNS
waveform (in blue).
}
 \label{fig:gw_convergence_and_surrogate}
\end{figure*}

\section{Numerical results}
\label{sec:numerical_results}
We present our main simulation results in this section. For comparison, the
BHNS system is evolved in both GR and ST theory, and two numerical
resolutions are adopted for each case by specifying different numerical error
tolerances to the adaptive mesh refinement (AMR) algorithm in SpEC
\cite{Szilagyi:2014fna}. Below we first give a qualitative panorama view of the
GR system in Sec.~\ref{subsec:bhns_gr}, and the ST system in
Sec.~\ref{subsec:bhns_st}. Then in Sec.~\ref{subsec:GR_vs_ST} we compare the GR and ST simulations. Finally in Sec.~\ref{subsec:bhns_pn}, we conduct more
quantitative discussions by comparing our numerical waveforms to existing
PN predictions in ST.

\subsection{The BHNS system in GR}
\label{subsec:bhns_gr}
We first evolve the system with GR, whose initial data are built based on the
method in Refs.~\cite{Foucart:2008qt,Tacik:2016zal}. For the GW200115-like binary parameters we consider (see Table~\ref{table:ns_properties}), the NS is
swallowed quickly by the BH during the merger, and there is no tidal
disruption. The remnant BH has a mass of $m_f=0.9785M$, with $M=7.2M_\odot$ the total Jordan-frame mass defined in Sec.~\ref{subsec:binary_parameter}. The remnant dimensionless
spin is $\chi_f=0.38$. As a standard numerical diagnostic, we plot the
volume-weighted generalized harmonic constraint energy [see Eq.~(53) of
Ref.~\cite{Lindblom:2005qh}] in Fig.~\ref{fig:constraint_gr},
where the orange (blue) curve refers to the low (high) resolution run. As
expected, the constraint energy decreases with increasing resolution, once the initial transients (known as junk radiation)
leave the domain $(t>R_{\rm bdry}=500M)$. Here $R_{\rm bdry}$ is the radius of our simulation domain, as summarized in Table~\ref{table:ns_properties}. In addition, we remark that the constraints jump drastically near $t=1938M$, when the NS starts to plunge into the BH.

The top panel of Fig.~\ref{fig:gw_convergence_and_surrogate} shows the dominant $l=m=2$ harmonic
$\tilde{h}_{22}$ emitted by the BHNS system, with low (in orange) and high (in
blue) resolution. We see that the two waveforms manifest significant dephasing near the
merger. Our current waveforms are less accurate than other recent BHNS SpEC simulations~\cite{Foucart:2020xkt} even though we use the same criteria to set the
    numerical error tolerances in AMR. This is mainly because the NS we consider is softer, which has a smaller radius and would require finer grids to resolve its structure. However, as the main purpose of this study is to get a first qualitative
understanding of BHNS binaries in ST, we expect the current accuracy to be
sufficient (see more details in Sec.~\ref{subsec:GR_vs_ST}).

The leading tidal effect in the GW phase evolution appears at 5PN order \cite{Flanagan:2007ix}, and is captured by a mass-weighted tidal parameter $\tilde{\Lambda}_2^{\rm GR}$ \cite{Wade:2014vqa}
\begin{align}
    \tilde{\Lambda}_2^{\rm GR}=\frac{16}{13}\frac{(M+11m^{\rm J}_{\rm BH})}{M^5}m^{\rm J\,4}_{\rm NS}\Lambda_2^{\rm GR}.
\end{align}
After plugging in the values listed in Table~\ref{table:ns_properties}, we find $\tilde{\Lambda}_2^{\rm GR}$ is around 2.95, implying that the emitted GWs are almost indistinguishable
from that of a BBH system with the same spins and mass ratio. To
demonstrate this, we compare the BHNS waveform to that of an equivalent BBH system (black dashed line in the top panel of Fig.~\ref{fig:gw_convergence_and_surrogate}). The data of the BBH binary are obtained from the
\texttt{NRSur7dq4} surrogate model \cite{Varma:2019csw}.
 We align the two waveforms $\tilde{h}_{22}^{\rm BHNS}$ and $\tilde{h}_{22}^{\rm Sur}$ by minimizing their mismatch $\mathcal{M}$:
\begin{align}
    \mathcal{M}=1-\frac{(\tilde{h}_{22}^{\rm BHNS}|\tilde{h}_{22}^{\rm Sur})}{\sqrt{(\tilde{h}_{22}^{\rm BHNS}|\tilde{h}_{22}^{\rm BHNS})(\tilde{h}_{22}^{\rm Sur}|\tilde{h}_{22}^{\rm Sur})}}, \label{eq:mismatch_def}
\end{align}
over time and phase shifts.
Here the time-domain inner product between two signals $a,b$ is given by
\begin{align}
  (a|b)={\rm Re}\,\int_{t_1}^{t_2}a(t)^{*}\ b(t) \ dt, \label{mismatch}
\end{align}
where the star denotes complex conjugation,
and we choose the optimization window to be $[t_1,t_2]=[200M,800M]$.
We provide the phase evolution $\phi_{22}$ of the aligned waveforms:
\begin{align}
    \phi_{22}\equiv {\rm{arg}}\,\tilde{h}_{22}, \label{eq:gw_phase_def}
\end{align}
in the middle panel of Fig.~\ref{fig:gw_convergence_and_surrogate}, as
well as the corresponding waveform phase differences $\Delta
\phi_{22}$ in the bottom panel. We see the phase difference between
the BHNS and BBH $(\sim 0.4\,{\rm rad})$ remains comparable to NR
numerical resolution difference up to $\sim 10M$ prior to the waveform peak, which
indicates that the tidal effect of this system is negligible.

\subsection{The BHNS system in ST: Scalar Field}
\label{subsec:bhns_st}
\begin{figure}[tb]
        \includegraphics[width=\columnwidth,clip=true]{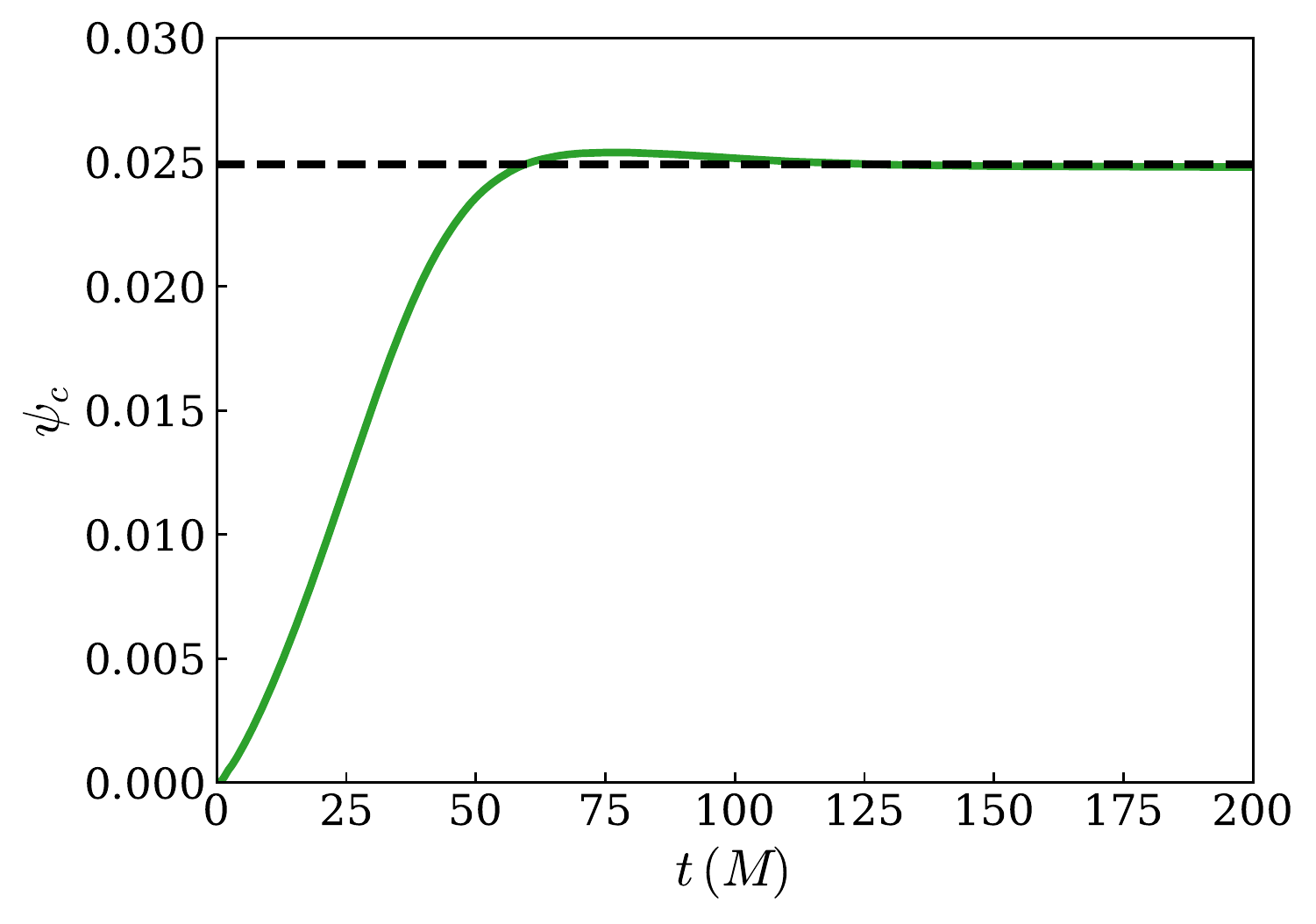}
\caption{The evolution of the scalar field $\psi$ measured at the center of the
NS. The plot describes the growth of the scalar field around the NS at the
beginning of the simulation. The horizontal dashed line corresponds to the
prediction by solving equations of motion for an isolated NS in
Sec.~\ref{subsec:scalar-tensor}.
}
 \label{fig:grow_psi}
\end{figure}

Let us then move on to the ST simulation. For simplicity, we use the same initial
data as its GR counterpart to evolve the system, where the scalar field is
absent;\footnote{It is straightforward to check that the GR initial data satisfy the ST constraint equations.}
while this means the initial data do not correctly capture a snapshot of the binary system in ST gravity that started at an infinite time in the past. This is also true for the GR simulation presented in Sec.~\ref{subsec:bhns_gr}, where Fig.~\ref{fig:constraint_gr} displays the presence of spurious initial transients during $t<R_{\rm bdry}=500M$. In our ST simulations, the system undergoes an extra transient regime
at the beginning of the evolution, during which a scalar field cloud grows
dynamically around the NS. In Fig.~\ref{fig:grow_psi}, we plot the
scalar field value $\psi_c$ measured at the stellar center as a function
of time. During the first $50M$, the value of $\psi_c$ increases and asymptotes to
the value predicted by the isolated NS solver (the horizontal dashed line) that
we used in Sec.~\ref{subsec:scalar-tensor}, which serves as a cross-check of
our numerical code. Note that the $\psi_c$ growth time scale is much shorter than the aforementioned initial transients $(t\sim500M)$, therefore we expect that
our results are not impacted by this additional transition from GR to ST.

\begin{figure}[tb]
        \includegraphics[width=\columnwidth,clip=true]{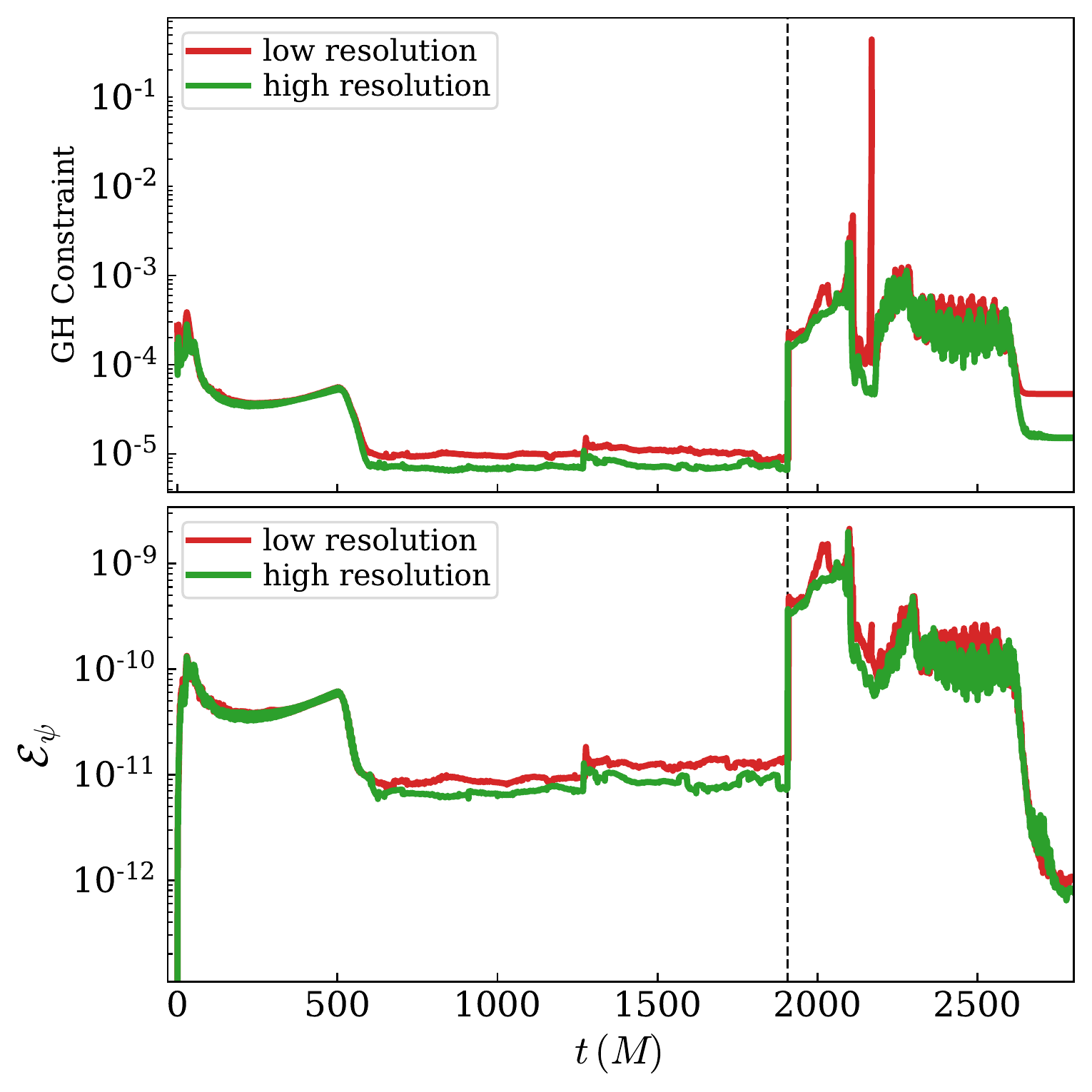}
\caption{
The evolution of the volume-weighted constraint energy for the metric (the
upper panel) and the scalar field (the lower panel), evolved with ST. The red (green) curve
corresponds to the low (high) resolution. The vertical dashed line indicates the onset of the merger.
}
\label{fig:constraint_st}
\end{figure}

We also provide the volume-weighted generalized harmonic
constraint energy [see Eq.~(53) of
Ref.~\cite{Lindblom:2005qh}] in the top panel of Fig.~\ref{fig:constraint_st} and find that the additional scalar field does not worsen the constraint violation compared to the GR system (Fig.~\ref{fig:constraint_gr}): the evolution of the constraint
is identical modulo a shift to an earlier time, due to the hastened merger of
the ST system. In addition, as for the scalar field's FOSH system \cite{Holst:2004wt,Scheel:2003vs}, we need to introduce an auxiliary dynamical variable $\Phi_i\equiv\partial_i\psi$, and its associated constraint energy:
\begin{align}
    \mathcal{E}_\psi=\left\lVert\sqrt{\sum_{i=1}^3 \left[C^{(1)}_iC^{(1)}_i+C^{(2)}_iC^{(2)}_i\right]}\right\rVert,
\end{align}
where $\lVert\cdot\rVert$ denotes $L^2$ norm over the domain. The derivative constraint for $\psi$, $C^{(1)}_i$,
reads
\begin{align}
    C^{(1)}_i=(\partial_i\psi)_{\rm num}-\Phi_i,
\end{align}
where $(\partial_i\psi)_{\rm num}$ corresponds to the numerical spatial derivative of $\psi$. The second derivative constraint for
$\psi$, $C^{(2)}_i$, is given by
\begin{align}
    C^{(2)}_i&=[ijk]\partial_j\Phi_k & (\text{sum on }j,k)
\end{align}
with $[ijk]$ being the Levi-Civita symbol, with $[123]=+1$. We provide the evolution of
$\mathcal{E}_\psi$ in the lower panel of Fig.~\ref{fig:constraint_st}. As expected, it also
decreases with increasing resolution.

\begin{figure*}[t]
    \centering
    \subfloat[The $x-y$ plane]{\includegraphics[width=0.779\columnwidth]{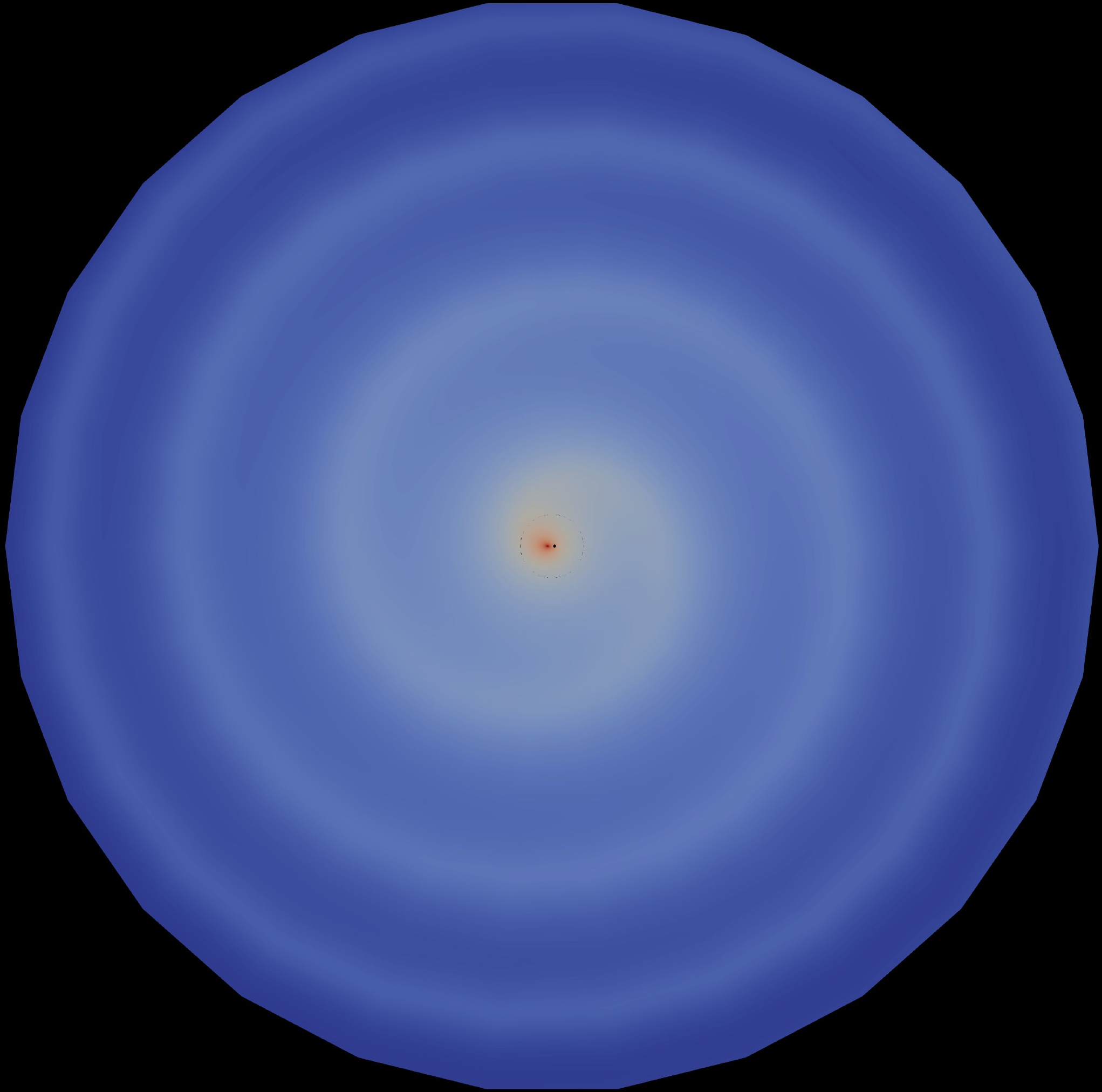}}
    \subfloat[The $y-z$ plane]{\includegraphics[width=0.8\columnwidth]{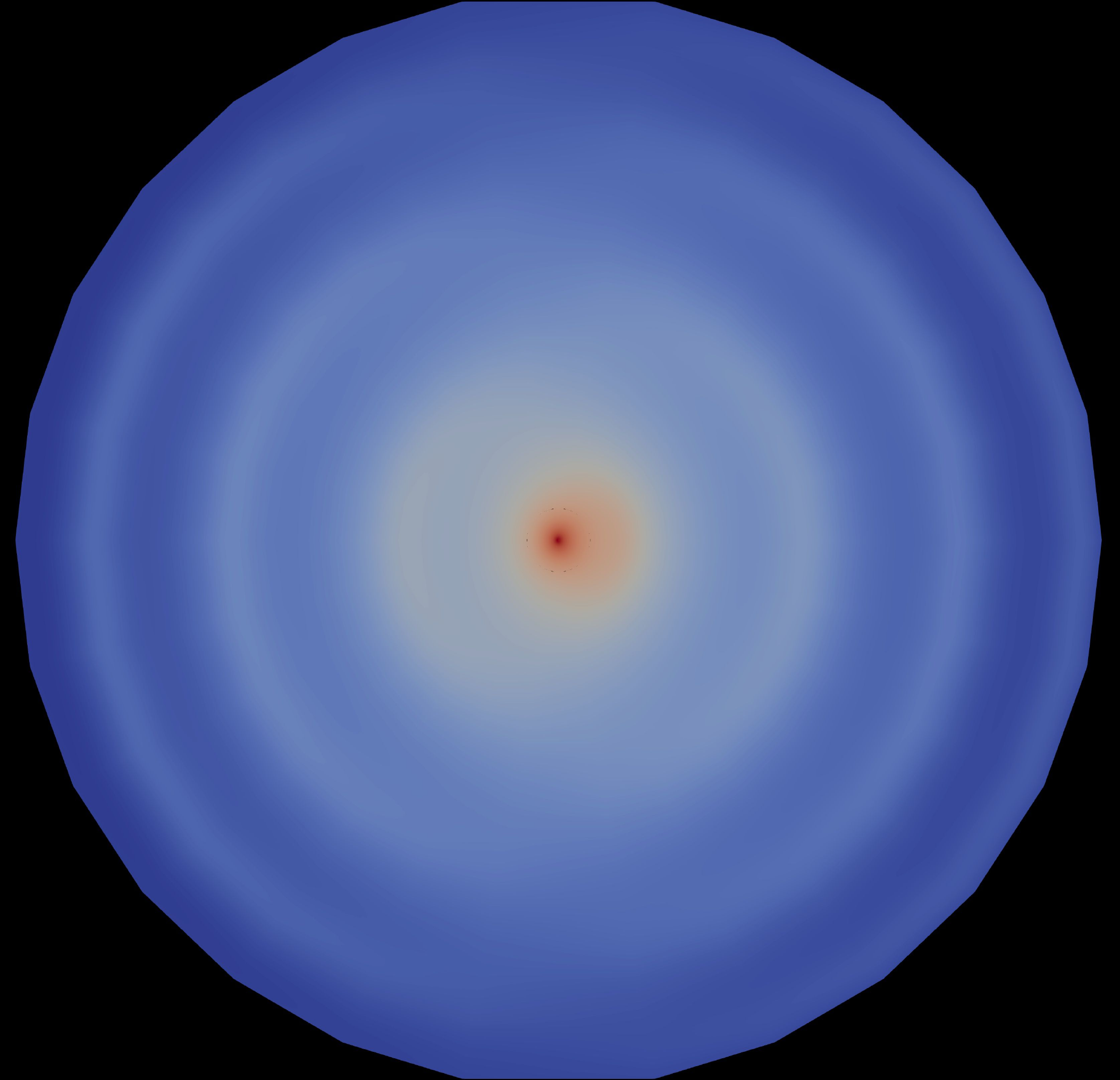}}
\caption{A snapshot of the field $\log|\psi|$ at $t=2062.3M$ across the entire
computational domain, with the outer boundary being at $500M$. The orbital
angular momentum is aligned with the $z$ axis.}
 \label{fig:scalar_field_visualization}
\end{figure*}

Finally, to close this subsection, we give a qualitative description of the scalar field $\psi$
in Fig.~\ref{fig:scalar_field_visualization} by taking a snapshot of its
distribution at $t=2062.3M$ across the entire computational domain. In
the wave zone, the distribution of the scalar field in the $x-y$ plane
(left panel) is singly periodic in $\varphi$
like $e^{i\varphi}$, where $\varphi$ is the azimuthal angle
defined in Eq.~\eqref{psi_extrapolate}; and in the $y-z$ plane (right
panel), we see vanishing on the $z$ axis with a single maximum at the
equatorial plane ($z=0$), like $\sin\iota$.  These patterns are
consistent with the dipolar nature $Y_{11}\sim\sin\iota
e^{i\varphi}$ of the scalar field, and we will discuss this in more
detail in Sec.~\ref{subsec:bhns_pn}.

\subsection{Comparison between the GR and ST}
\label{subsec:GR_vs_ST}

\begin{figure}[tb]
        \includegraphics[width=\columnwidth,clip=true]{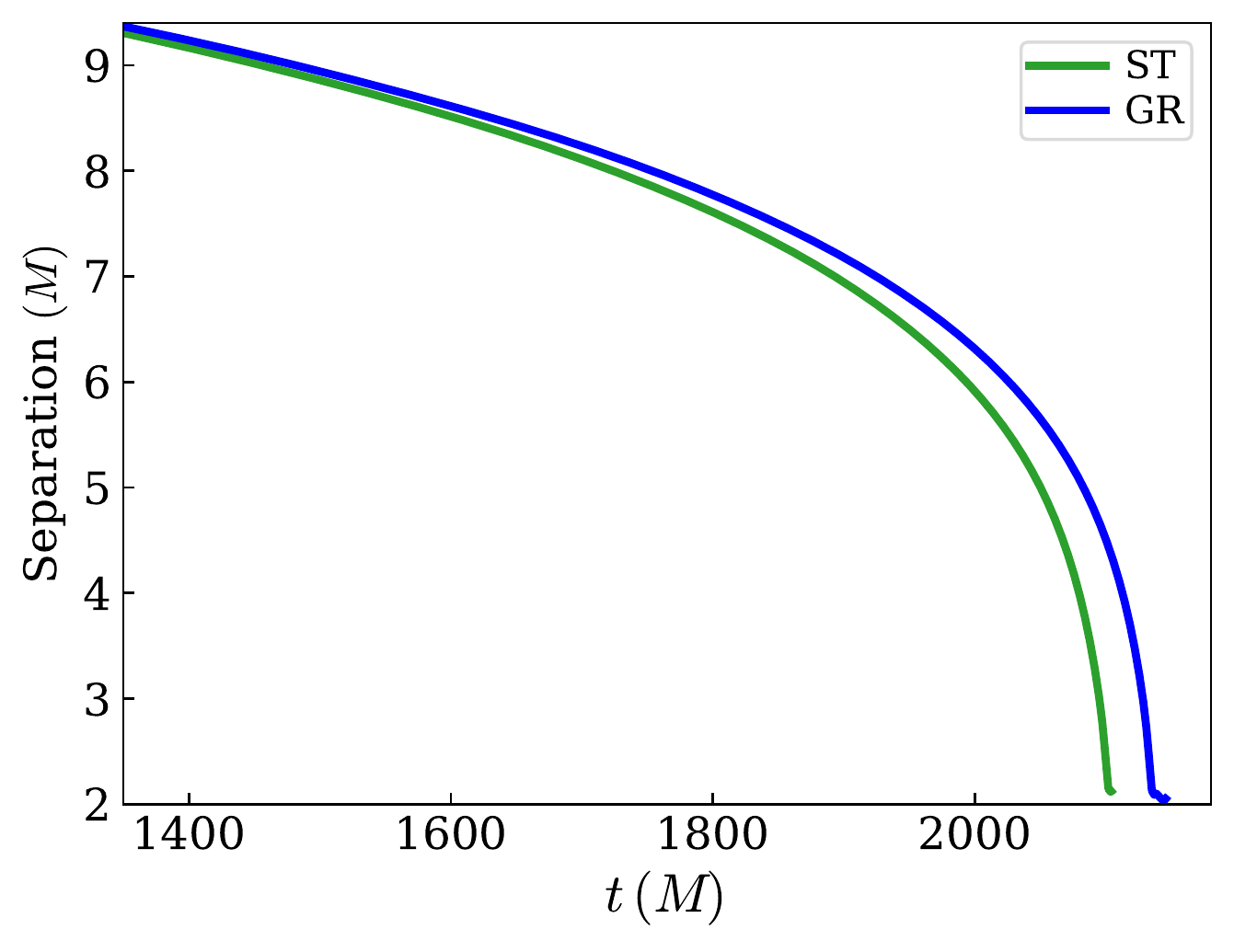}
\caption{The evolution of the orbital separation for the BHNS system, in the ST
    gravity (green) and GR (blue).
  }
 \label{fig:orbital_separation_st_vs_gr}
\end{figure}

Figure \ref{fig:orbital_separation_st_vs_gr} displays the evolution of the coordinate separation between the two compact objects for the GR and the ST systems. We first see that the merger portions of both systems can be
aligned perfectly through a time shift, namely, they have a similar $\dot{R}-R$ dependence near the merger and thus a similar plunge dynamic, implying a similar orbital separation (and therefore similar frequency) for the onset of the plunge. This feature is different from the
BNS simulations in Ref.~\cite{Barausse:2012da}, where ST binaries were found to
merge at significantly larger orbital separation (see their Fig.~1). The
difference arises from the size of the gravitational attraction. Recall that the
gravitational pull in ST gravity is characterized by the effective gravitational
constant $G_{\rm eff}=G_*(1+\alpha_{\color{red}{A}}\alpha_{\color{red}{B}})$ \cite{Damour:1992we}, which is
amplified for BNS systems when both the NSs have a nonzero scalar charge. Consequently, their
plunges happen at larger orbital separations. By contrast, the gravitational pull
in our ST BHNS system is similar to its GR counterpart because the BH's scalar
charge vanishes, so the scalar sector has negligible impact on
the plunge separation. 
However, the ST simulation does exhibit a nonnegligible deviation from its GR counterpart over a longer timescale. As shown in Fig.~\ref{fig:orbital_separation_st_vs_gr}, the ST
simulation has a shorter total duration than the GR case, even though they both start at the same separation. This is because the
scalarized NS admits an additional energy dissipation channel via scalar
radiation; therefore the system in ST gravity evolves faster during the
inspiral. 

\begin{figure*}[tb]
        \includegraphics[width=\textwidth,clip=true]{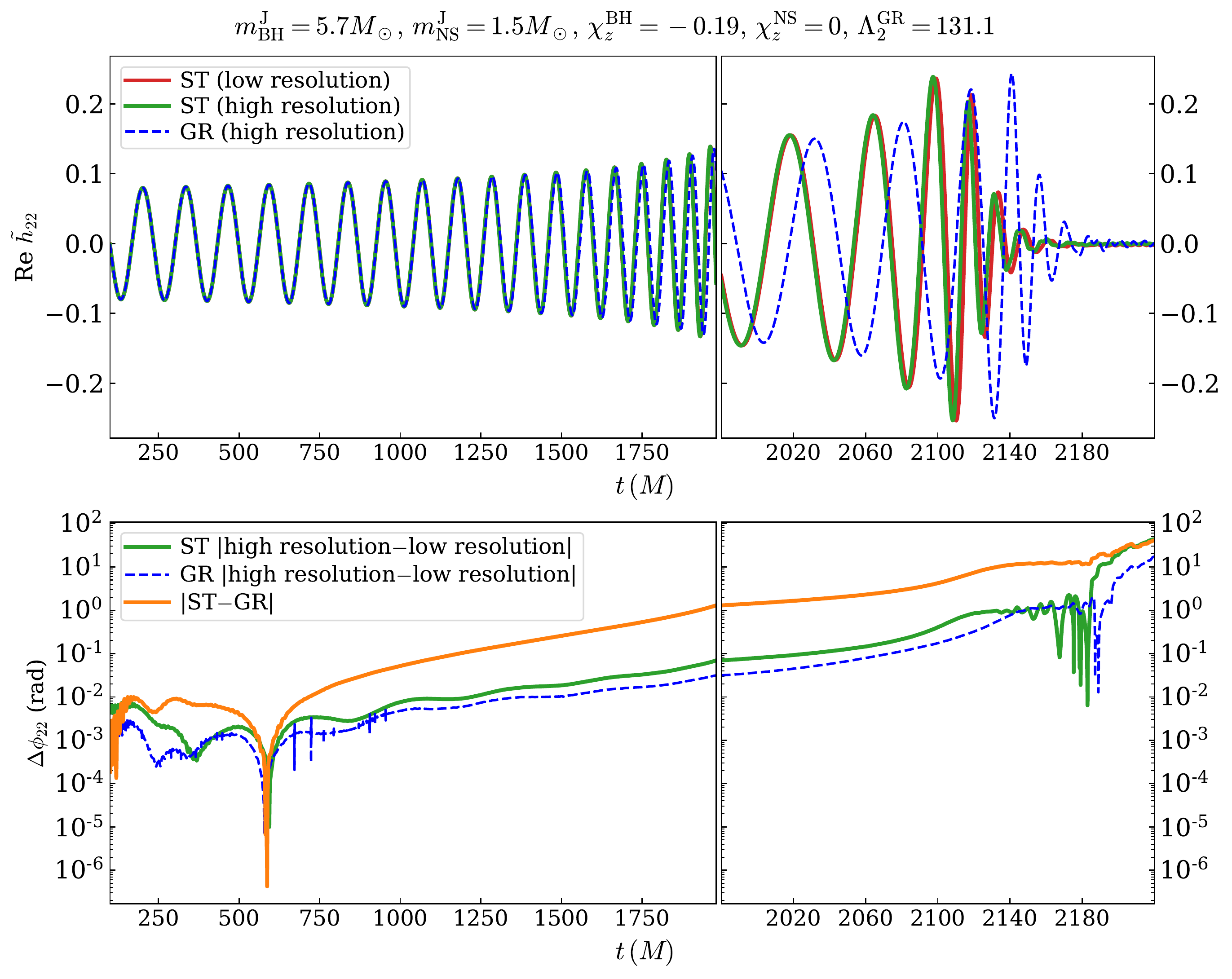}
\caption{Upper panel: The ST waveforms with a low (in red) and high (in green) resolution. They are
compared to the GR waveform (in blue). Lower panel: the phase difference
between the GR and ST waveforms (in orange). For reference, the numerical resolution differences of the GR
and the ST waveform are also presented in blue and green, respectively. In addition,
we summarize some of the binary parameters in the title.
}
 \label{fig:GR-ST}
\end{figure*}

A direct consequence of the hastened dynamics is a shortening of the GW signal. Figure \ref{fig:GR-ST} provides the $l=m=2$
harmonic of the ST waveform for two different resolutions (solid curves). For
reference, $\tilde{h}_{22}$ in GR is plotted as the blue dashed curve. 
Here we still align the waveforms by minimizing the mismatch in Eq.~\eqref{eq:mismatch_def} over time and phase shifts. The same time window $[t_1,t_2]=[200M,800M]$ is used.
After the
peak of the ST waveform,
it takes the GR waveform an extra GW cycle, $\Delta \phi_{22}\sim 6.34\, {\rm rad}$ [Eq.~\eqref{eq:gw_phase_def}], to reach its peak, smaller than GR's numerical resolution difference at the peak $(\sim 0.6\, {\rm rad})$. Therefore our simulations are able to capture the effect of scalar radiation well above the numerical resolution difference, even though our simulations are less accurate than other recent BHNS SpEC simulations~\cite{Foucart:2020xkt}, as discussed in Sec.~\ref{subsec:bhns_gr}.

\subsection{Comparing to post-Newtonian theory}
\label{subsec:bhns_pn}
We now carry out quantitative comparisons between the simulated GW waveforms and existing PN waveform predictions in ST. 
As pointed out in
Refs.~\cite{Sennett:2016klh,Bernard:2022noq}, the relative size of the leading
scalar dipolar radiation and leading tensor quadrupolar radiation is given by
\begin{align}
   \frac{\mathcal{F}_{\rm nd}}{\mathcal{F}_{\rm d}} =\left(\frac{24}{5\zeta\mathcal{S}_-^2}\right)x,
\end{align}
with $\mathcal{F}$ being energy flux.  In our simulation, we find the factor above is greater than 25,
i.e. quadrupolar radiation dominates, so we are in the quadrupole-driven
regime~\cite{Sennett:2016klh}.

\begin{figure*}[tb]
        \subfloat[ST\label{fig:GR_PN_ST}]{        \includegraphics[width=\textwidth,clip=true]{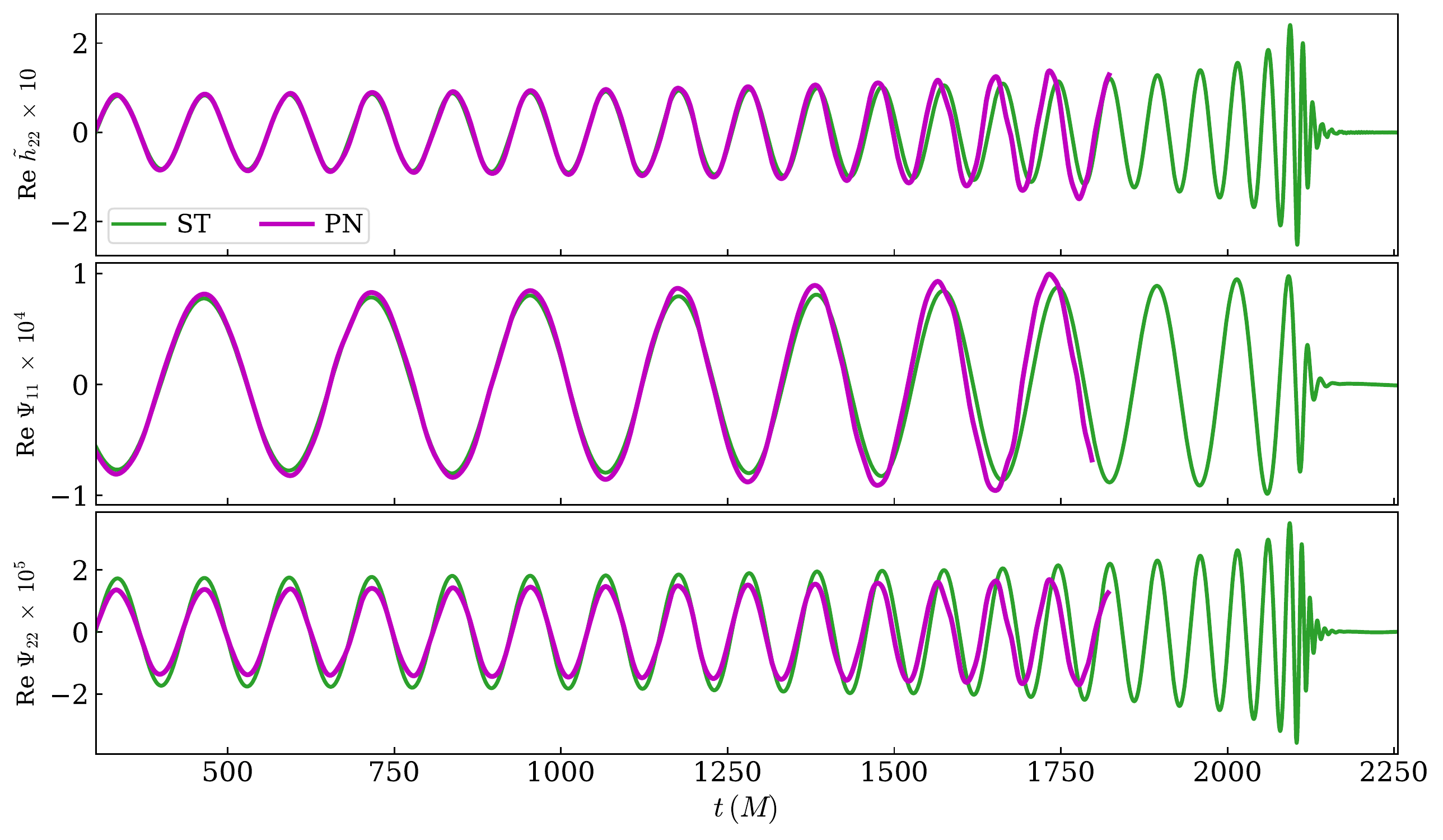}}\\
        \subfloat[GR\label{fig:GR_PN_GR}]{\includegraphics[width=\textwidth,clip=true]{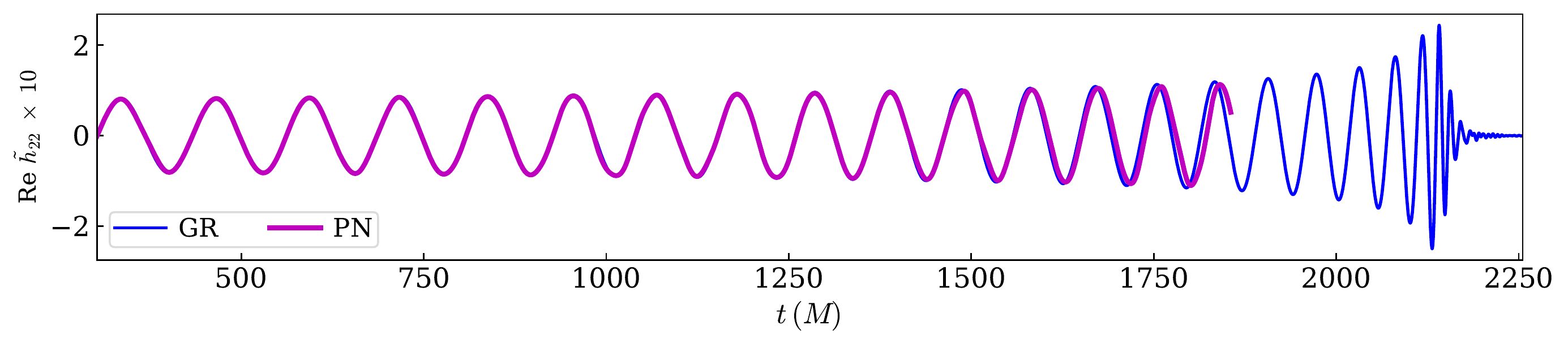}}
\caption{Comparing the numerical waveforms (in green and blue) to the PN model
(in magenta). Fig.~\ref{fig:GR_PN_ST} shows the ST tensor harmonic
$\tilde{h}_{22}$ (top) and the scalar modes $\Psi_{11}$ (middle) and
$\Psi_{22}$ (bottom). Note that the modes $\Psi_{lm}$ are defined in Eq.~\eqref{eq:Psilm_psilm}. Fig.~\ref{fig:GR_PN_GR} provides the GR tensor harmonic
$\tilde{h}_{22}$.
}
 \label{fig:GR_PN}
\end{figure*}

We first consider the gravitational modes
$\tilde{h}_{lm}$, whose PN expressions read \cite{Sennett:2016klh}
\begin{align}
    \tilde{h}_{lm}=2\tilde{G}(1-\zeta)\eta x\sqrt{\frac{16\pi}{5}}\hat{H}_{lm}e^{-im\phi}, \label{hlm-GW}
\end{align}
where $\eta=m^{\rm J}_{\rm BH}m^{\rm J}_{\rm NS}/(m^{\rm J}_{\rm BH}+m^{\rm
J}_{\rm NS})^2$ is the symmetric mass ratio, $x=(\tilde{G}M\alpha \Omega_{\rm
orb})^{2/3}$ is the PN expansion parameter, $\Omega_{\rm orb}$ is the
orbital frequency, and we give $\phi$ below.  We summarize the definition of ST parameters
$\tilde{G},\zeta,\alpha$ in Table~\ref{table:ST_PN}.
In Eq.~\eqref{hlm-GW}, comparing with Eq.~(65) of Ref.~\cite{Sennett:2016klh},
we removed an overall factor $M/r$ which is already divided out in
Eq.~\eqref{h22_extrapolate}. The expressions for $\hat{H}_{lm}$ are long and
they can be found in Eqs.~(67) of Ref.~\cite{Sennett:2016klh}.
Because the dipolar scalar radiation starts 1PN earlier than the
leading quadrupolar gravitational radiation,
the inspiral is separated into two parts: dipolar (D) or
non-dipolar (ND).
The phase factor $\phi$ reads
\begin{subequations}
  \begin{align}
    \label{PN_phase_total}
    \phi &= \phi_{\rm nd} + \phi_{\rm d} \,,\\
    \label{PN_phase_nd}
    \phi_{\rm nd} &=-\frac{x^{-5/2}}{32\eta\xi}\left[1+\frac{5}{3}\rho_2^{\rm nd}x+\frac{5}{2}\rho_3^{\rm nd}x^{3/2}+5\rho_4^{\rm nd}x^2\right. \notag \\
    &\qquad\left. +\frac{5}{2}\rho_3^{\rm spin}x^{3/2}+5\rho_4^{\rm spin}x^2\right] \,, \\
    \label{PN_phase_d}
    \phi_{\rm d}&=\frac{25\mathcal{S}_-^2\zeta x^{-7/2}}{5376\eta\xi^2}\left[1+\frac{7}{5}\rho_2^{\rm d}x+\frac{7}{4}\rho_3^{\rm d}x^{3/2}+\frac{7}{3}\rho_4^{\rm d}x^2\right]
    \,,
  \end{align}
\end{subequations}
with the coefficients $\rho_{i}^{\rm nd/d}$'s being listed in Eqs.~(B10) of
Ref.~\cite{Sennett:2016klh}. The ST parameters $\xi$ and
$\mathcal{S}_\pm$ are defined in Table~\ref{table:ST_PN}, and we see that all of them depend on the sensitivity of the NS
\begin{align}
    s_{\rm NS}=\left(\frac{d \ln m^{\rm J}_{\rm NS}}{d \ln\phi}\right)_{\phi_0}.
\end{align}
The relationship between $s_{\rm NS}$ and the scalar charge $\alpha_{\rm NS}$
reads \cite{Sennett:2016klh}
\begin{align}
  \label{eq:sensitivity_charge}
  s_{\rm NS}=\frac{1}{2}-\frac{\alpha_{\rm NS}}{2\alpha_0}
  \,,
\end{align}
where $\alpha_0$ is the ST parameter defined in Eq.~\eqref{omega_psi}.
Equation~\eqref{PN_phase_nd} is controlled by the quadrupolar
radiation, while Eq.~\eqref{PN_phase_d} is controlled by the dipolar radiation starting at
$-1$PN. Spin effects are not considered in Ref.~\cite{Sennett:2016klh};
here we simply add the spin contributions in GR, leading to the second line
in Eq.~\eqref{PN_phase_nd}, and we leave the relevant ST corrections for
future studies. The expressions of $\rho_{i}^{\rm spin}$'s can be found in
Eq.~(4.16) of Ref.~\cite{Kidder:1995zr},
\begin{align}
    &\rho_3^{\rm spin}=\frac{1}{12}\sum_{i=1,2}\chi_i(\hat{\bm{L}}_N\cdot\hat{\bm{s}}_i)\left(113\frac{m_i^2}{M^2}+75\eta\right), \\
    &\rho_4^{\rm spin}=\frac{1}{48}\eta\chi_1\chi_2[247(\hat{\bm{s}}_1\cdot\hat{\bm{s}}_2)-721(\hat{\bm{L}}_N\cdot\hat{\bm{s}}_1)(\hat{\bm{L}}_N\cdot\hat{\bm{s}}_2)],
\end{align}
where $\hat{\bm{L}}_N$ and $\hat{\bm{s}}_i$ stand for the unit vector along the
orbital angular momentum and the individual spin $\bm{s}_i$. Furthermore, we
note that tidal effects are ignored in Eq.~\eqref{PN_phase_total}, which
formally enter into the phase evolution at 5PN order~\cite{Flanagan:2007ix}. This is reasonable for this study, as the system's mass-weighted tidal deformability $\tilde{\Lambda}_2^{\rm GR}\sim 2.95$ is very small and it has little impact
on the binary dynamics, as shown in Fig.~\ref{fig:gw_convergence_and_surrogate}. In the top panel of
Fig.~\ref{fig:GR_PN_ST}, we compare the ST numerical waveform $\tilde{h}_{22}$ to the PN prediction, finding good agreement until
$\sim500M$ before the merger. For reference, we also plot the GR waveform
$\tilde{h}_{22}$ and the corresponding PN prediction in Fig.~\ref{fig:GR_PN_GR}. Additionally, in App.~\ref{app:Hierarchical}, we present a more detailed comparison by demonstrating the hierarchical contributions of each PN term.

\begin{table}
    \centering
\renewcommand\arraystretch{1.6}
    \caption{Summary of PN parameters used for ST gravity. We have used the fact
    that a BH's scalar charge vanishes: $\alpha_{\rm BH}=0$, and thus $s_{\rm BH}=1/2$ following Eq.~\eqref{eq:sensitivity_charge}. Note that $\alpha$ is not to be confused with the scalar
charge $\alpha_{\rm NS}$.}
\begin{tabular}{c@{\hspace{0cm}}c@{\hspace{0cm}}c@{\hspace{0cm}}c@{\hspace{0cm}}c@{\hspace{0cm}}c@{\hspace{0cm}}c}
\hline\hline
$\omega_0$ & $\tilde{G}$ & $\zeta$ & $\alpha$ & $\mathcal{S}_-$ & $\mathcal{S}_+$ & $\xi$ \\
\hline
\cellcolor{lightgray} $\frac{1-3\alpha_0^2}{2\alpha_0^2}$ & \cellcolor{lightgray} $\frac{1+\alpha_0^2}{\phi_0}$ & \cellcolor{lightgray} $\frac{\alpha_0^2}{1+\alpha_0^2}$ & \cellcolor{lightgray} $\frac{1}{1+\alpha_0^2}$ & \cellcolor{lightgray} $-\alpha^{1/2}s_{\rm NS}$ & \cellcolor{lightgray} $\alpha^{1/2}(1-s_{\rm NS})$ & \cellcolor{lightgray} $1+\frac{\zeta \mathcal{S}_+^2}{6}$ \\
\end{tabular}
     \label{table:ST_PN}
\end{table}

We then compare the scalar modes $\psi_{l m}$ extracted from our simulation with predictions from PN. The PN prediction for the $(l,m)$ harmonic of the transverse breathing mode $\Psi$
[see Eq.~\eqref{eq:metric_scalar_tensor_decomp}] is given by \cite{Bernard:2022noq}
\begin{align}
    \Psi_{lm}=2i\tilde{G}\zeta\sqrt{\alpha}\mathcal{S}_-\eta\sqrt{x}\sqrt{\frac{8\pi}{3}}\hat{\Phi}_{lm}e^{-im\phi}, \label{scalar_mode_pn}
\end{align}
where the expression of $\hat{\Phi}_{lm}$ can be found in Eqs.~(6.10) of
Ref.~\cite{Bernard:2022noq}; and $\Psi_{lm}$ is defined in parallel with
Eq.~\eqref{psi_extrapolate}:
\begin{align}
    r\Psi/M =\sum_{l,m}Y_{lm}(\iota,\varphi)\Psi_{lm}.
\end{align}
Here $\Psi_{lm}$ is related to our numerical extracted scalar mode $\psi_{lm}$
[Eq.~\eqref{psi_extrapolate}] via
\begin{align}
    \Psi_{lm}=-4\sqrt{\pi}\alpha_0\psi_{lm}, 
    \label{eq:Psilm_psilm}
\end{align}
where Eq.~\eqref{phi_psi} has been used. 
We compare our numerical scalar
modes $\Psi_{11}$ and $\Psi_{22}$ to the PN predictions in the middle and
bottom rows of Fig.~\ref{fig:GR_PN_ST}, and refer to
App.~\ref{app:other_modes} for other (subdominant) modes. Similar to
$\tilde{h}_{22}$, the PN predictions for the $\psi_{l m}$ phase evolution are accurate until
$\sim500M$ before merger; however, their amplitudes do not match as accurately
as their phases.

\section{Waveform distinguishability}
\label{sec:parameter_estimation_bias}
We have discussed features of the BHNSs in GR and ST. Then in this
section, we investigate how our numerical simulations can help place
constraints on ST theory with GW200115 and future BHNS observations. Specifically, here we focus on whether a ST waveform can be distinguished from a GR waveform. We estimate this by computing the mismatch $\mathcal{M}$ between the two
waveforms, defined in Eq.~\eqref{eq:mismatch_def}. Note that in
Eq.~\eqref{mismatch}, we used a flat noise curve for simplicity, namely assuming an idealized detector.

\begin{figure}[tb]
        \includegraphics[width=\columnwidth,clip=true]{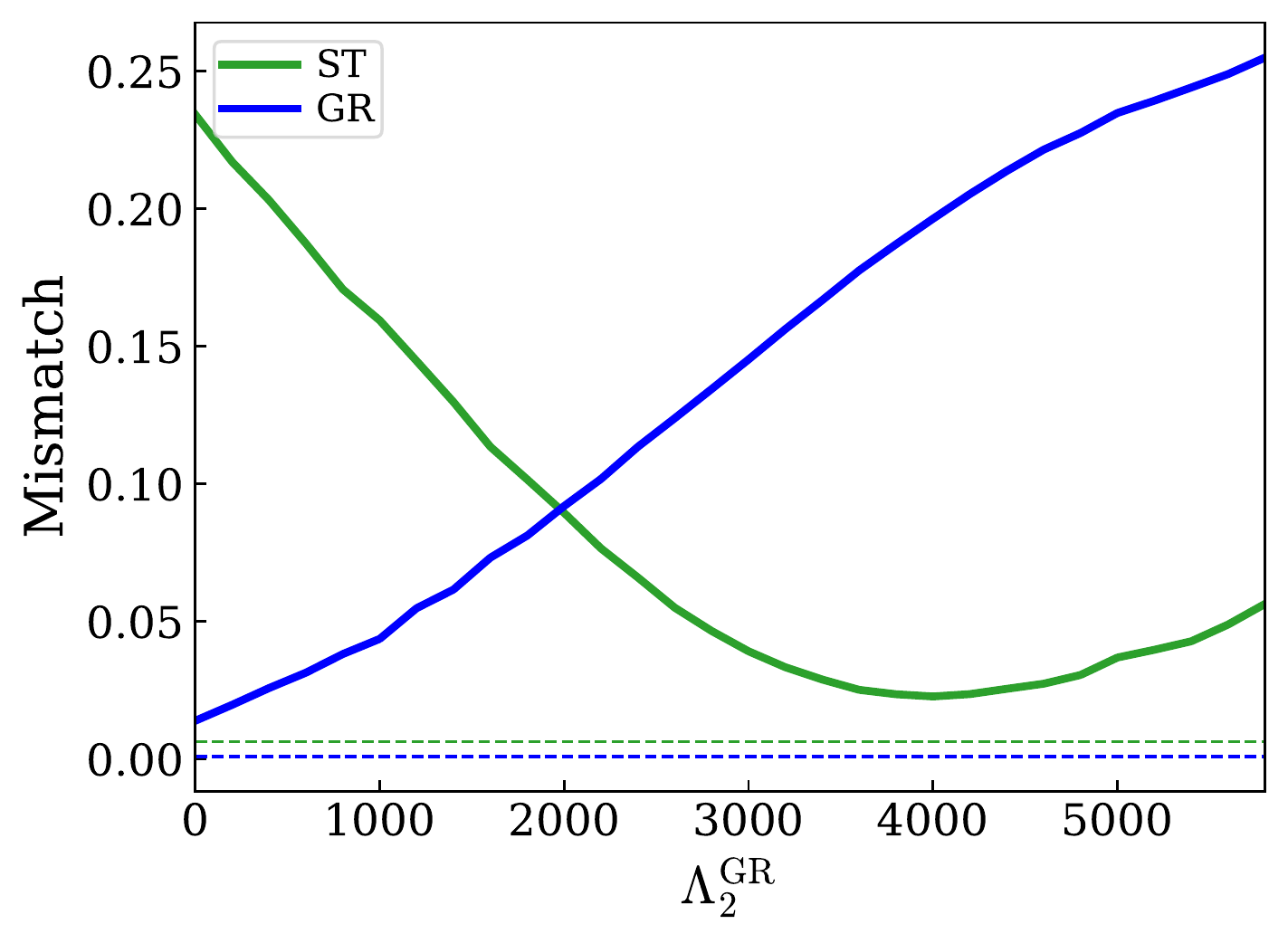}
\caption{The mismatch of the \texttt{SEOBNRv4T} model with the ST waveform
(green) and the GR result (blue), as a function of tidal deformability
$\Lambda_2^{\rm GR}$. For the sake of comparison, we also compute the mismatch between
two resolutions for ST (green dashed line) and GR (blue dashed line).}
 \label{fig:mismatch}
\end{figure}

\begin{figure*}[tb]
    \centering
    \includegraphics[width=2\columnwidth,clip=true]{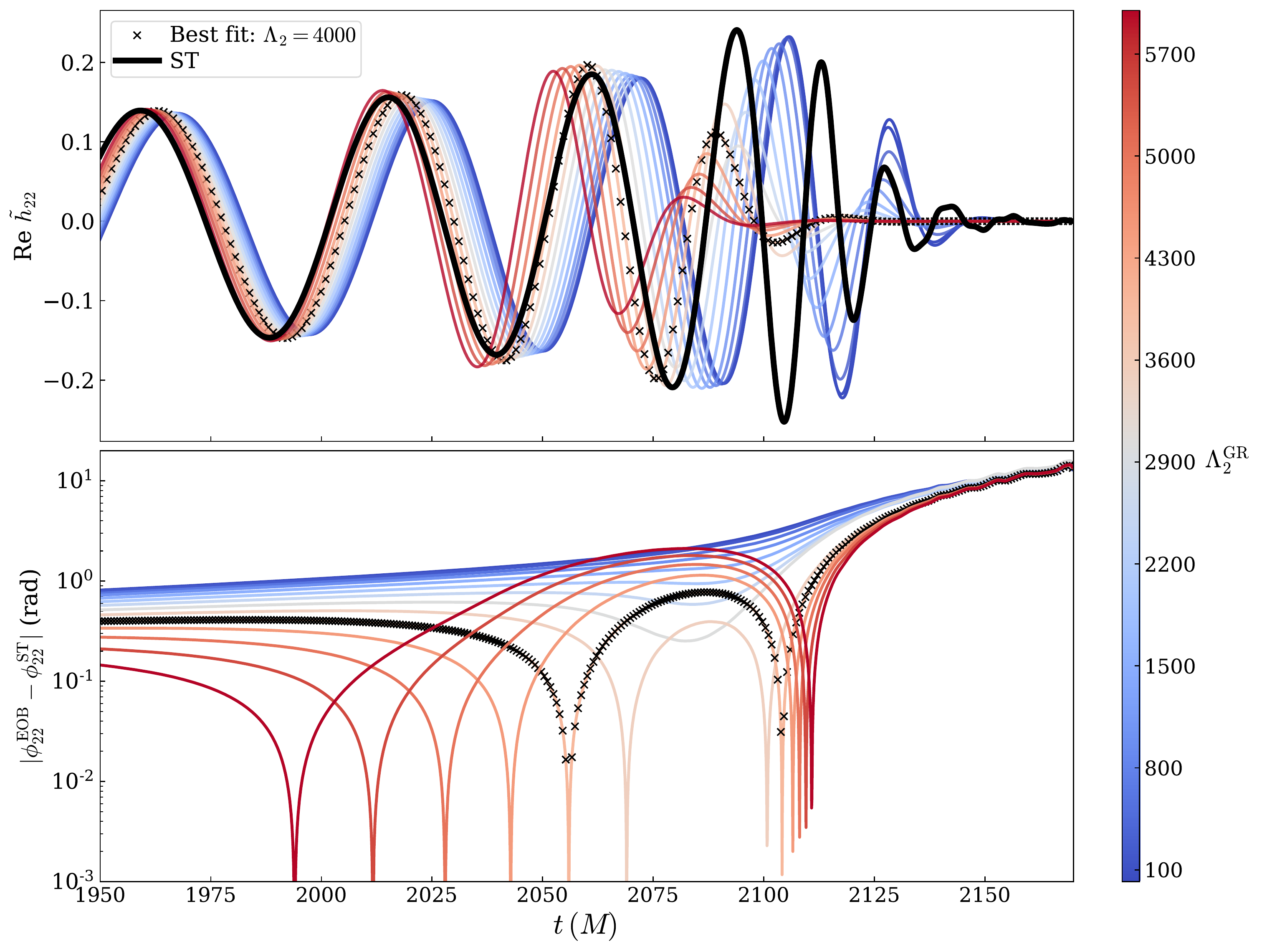}
\caption{Comparing the ST waveform (black) with the \texttt{SEOBNRv4T} model, with a variety of $\Lambda_2^{\rm GR}$, ranging from 0 to 6000. The minimum mismatch $\mathcal{M}\sim0.023$ happens at $\Lambda_2^{\rm GR}\sim4000$.
}
 \label{fig:EOB}
\end{figure*}

We first compute the mismatch between the GR and ST waveform $\tilde{h}_{22}$ presented in Fig.~\ref{fig:GR-ST} and find $\mathcal{M}=0.38$. Since the error in our simulations is larger than other BHNS SpEC simulations (see discussions around Figs.~\ref{fig:gw_convergence_and_surrogate} and \ref{fig:GR-ST}), we terminate the integration in Eq.~\eqref{mismatch}
at the peak of the ST waveform $(t_2=2102M)$ to avoid the ringdown region. One criterion for the distinguishability of two waveforms reads
\cite{Flanagan:1997kp,Lindblom:2008cm,McWilliams:2010eq,Chatziioannou:2017tdw,Boyle:2019kee}
\begin{align}
    \mathcal{M}>\frac{D}{2\rho^2}, \label{distinguishability}
\end{align}
where $D=5$ is the number of free intrinsic parameters (chirp mass, mass ratio, spin magnitudes on both compact objects, and tidal deformability) of our nonprecessing systems, and
$\rho$ is the signal-to-noise ratio (SNR). After inserting the numbers, we find $\rho>2.56$ is needed to
distinguish ST from GR. Such a low SNR threshold is not surprising for this specific case with extreme scalarization and an idealized detector, given the significant dephasing between the two waveforms shown in Fig.~\ref{fig:GR-ST}. For more moderate ST parameters and more realistic detectors, the deviation is not expected to be as large, and we leave this exploration for future work.

The subsequent question to consider is the extent to which tidal effects within GR can replicate the ST waveform. To explore this question, we employ an effective-one-body (EOB) model known as \texttt{SEOBNRv4T} \cite{Hinderer:2016eia,Steinhoff:2016rfi}. This model includes tidal effects and is characterized by tidal deformability coefficients $\Lambda_l$ in its tidal sector, with $l = 2$ being the focus in this case. To generate the \texttt{SEOBNRv4T} waveforms with varying $\Lambda_2^{\rm GR}$, we utilize LALSuite \cite{lalsuite}. Figure \ref{fig:mismatch} showcases the mismatch of these waveforms with the ST waveform $\tilde{h}_{22}$ as a function of $\Lambda_2^{\rm GR}$ while fixing other intrinsic parameters at their NR values. The mismatch first
decreases when $\Lambda_2^{\rm GR}$ is small, and the best match
$\mathcal{M}\sim 0.023$ happens at $\Lambda_2^{\rm GR}\sim 4000$. As a comparison, we repeat the same calculation for the mismatch between the \texttt{SEOBNRv4T} model and the GR waveform. The result is shown as the blue curve in Fig.~\ref{fig:mismatch}, and we can see the mismatch grows monotonically with
$\Lambda_2^{\rm GR}$ (recall the tidal effect is negligible in the GR
simulation).
To better understand the feature, in Fig.~\ref{fig:EOB} we provide the \texttt{SEOBNRv4T} waveforms with a variety of $\Lambda_2^{\rm GR}$, ranging from 0 to 6000. In particular, we mark the best-fit waveform ($\Lambda_2^{\rm GR}= 4000$) with black crosses. With increasing $\Lambda_2^{\rm GR}$, we see the tidal waveforms gradually shift backward in time, because the tidal effect accelerates the evolution and shortens the length of waveforms. This behavior is similar to the effect of the scalar field and dipole radiation. Notably, as $\Lambda_2^{\rm GR}$ approaches 4000, the last two wave cycles of the \texttt{SEOBNRv4T} waveforms (at $t\sim2075M$) align more closely with ST's phase evolution, resulting in a smaller mismatch. Further increasing $\Lambda_2^{\rm GR}$ beyond this point causes the tidal waveforms to deviate again from the ST waveform. Therefore, the mismatch in Fig.~\ref{fig:mismatch} bounces back.

Our preliminary mismatch comparison shows that both the
tidal and scalar sectors could produce similar and potentially degenerate imprints in GWs given the length of our simulations ($\sim12$ cycles before the merger). A limitation of our analysis is that the NR waveforms are relatively short and lacked low-frequency components --- the dipole radiation appears at $-1$PN whereas the tidal effect at 5PN. A longer waveform with a broader frequency span may break the degeneracy. A more comprehensive analysis is therefore necessary to fully characterize these features using longer waveforms with a broader frequency span and Bayesian parameter estimation. We leave this exploration to future research.

\section{Conclusion}
\label{sec:conclusion}
In this paper, we numerically simulate a fully relativistic BHNS binary system in ST theory, chosen to be consistent with GW200115 \cite{LIGOScientific:2021qlt}. To maximize the effect of spontaneous
scalarization, we set the ST parameters $(\beta_0, \alpha_0)$ to be at the boundary of known constraints from other observations \cite{Berti:2015itd}: $(-4.5,
-3.5\times10^{-3})$. In addition, we select a soft EOS for the NS so that it can generate a large scalar charge, as summarized in Table~\ref{table:ns_properties}. Following
Refs.~\cite{Foucart:2008qt,Tacik:2016zal}, we construct the initial data without including the scalar sector. Instead, the scalar field dynamically grows during the first $\sim50M$,
and quickly approaches the desired value predicted by the isolated NS solver.

We evolve the BHNS system with both GR and ST. For the GR binary, we find the soft EOS results in GW emissions that are nearly identical to
those of a BBH system with the same spins and mass ratio. In contrast, the ST
binary exhibits dominant dipolar radiation due to spontaneous
scalarization, with the spatial distribution of the scalar field $\psi$
matching the dipolar emission pattern throughout the computational domain. As a
result of this additional dipolar radiation, the ST binary evolves faster than
its GR counterpart, and the ST binary reaches its peak amplitude one whole GW cycle earlier than the GR counterpart.  We also compare our waveforms, including the tensor mode
$\tilde{h}_{22}$ and scalar breathing modes $\Psi_{11,22}$, with existing PN
waveform predictions in ST
\cite{Sennett:2016klh,Bernard:2022noq,Kidder:1995zr}, and find reasonable
agreement up to $\sim 500M$ before the merger. Finally, we compute the mismatch
between our ST waveform and the \texttt{SEOBNRv4T} model as a function of tidal
deformability $\Lambda_2^{\rm GR}$. We find the ST waveform could be partially mimicked
by a GR tidal waveform with a large $\Lambda_2^{\rm GR}\sim 4000$, due to the tidal
effect accelerating the evolution of the binary.

Throughout the analysis, we pick optimal choices for the EOS and the ST theory
parameters in order to produce a significant scalarization effect, and thus
strong dipolar radiation. Under this idealized scenario, we find that the GR and ST waveforms should be distinguishable for SNRs above 2.56. To fully understand observational prospects of constraining ST theory using BHNS systems, future work should explore a wider range of EOSs and
more moderate ST parameters. Specifically, the
scalar field's ability to alter the properties of NSs, such as compactness and
radius, may play a crucial role in determining whether the NSs are
disrupted or
not \cite{Foucart:2018rjc}, potentially leading to rich phenomena in the corresponding GW and even
electromagnetic emissions for ST binary systems.

Our mismatch tests using the \texttt{SEOBNRv4T} model and the GR waveforms indicate that the ST sector might be partially degenerate with tidal effects during the late inspiral stage (excluding low-frequency regime),
which can lead to parameter estimation biases. Here we restrict ourselves
to a single degree of freedom: $\Lambda_2^{\rm GR}$, while holding other parameters such
as mass ratio and spins constant. A possible avenue for future work is to carry
out a more systematic full Bayesian parameter estimation to better account for
these degeneracies.

Finally, our waveforms are obtained at null infinity through extrapolation
following Refs.~\cite{Iozzo:2020jcu,Boyle:2013nka,Boyle:2014ioa,Boyle:2015nqa}, with the $\texttt{PYTHON}$ package $\texttt{scri}$ \cite{scri,mike_boyle_2020_4041972}. The method is an approximate approach that
relies on the asymptotic behavior of several fields given by the peeling
theorem \cite{Newman:1961qr}. While this approximate approach captures linear
signals, it does not accurately capture nonlinear features such as the memory effect
\cite{Du:2016hww,Koyama:2020vfc,Hou:2020tnd,Seraj:2021qja,Mitman:2020pbt,Mitman:2020bjf}.
The more correct Cauchy-Characteristic Extraction (CCE)
\cite{Moxon:2020gha,Moxon:2021gbv} method would be required to fully account
for these effects. Therefore, another future avenue could be to evolve
the coupled metric-scalar system using a CCE framework adapted to ST,
and investigate the memory effect in ST
gravity~\cite{Du:2016hww, Koyama:2020vfc, Hou:2020tnd, Seraj:2021qja,
  Tahura:2020vsa, Tahura:2021hbk,Heisenberg:2023prj}.

\begin{acknowledgments}
We thank Laura Bernard, David Trestini, Luc Blanchet, Noah Sennet, Sylvain Marsat, and Alessandra Buonanno for sharing Mathematica notebooks with PN expressions.
We thank David Trestini, Hector Silva, and Dongze Sun for useful discussions.
V.V. acknowledges funding from the European Union's Horizon 2020 research and
innovation program under the Marie Skłodowska-Curie grant agreement No.~896869.
V.V. was supported by a Klarman Fellowship at Cornell.
L.C.S. was partially supported by NSF CAREER Award
PHY-2047382. S.M. and M.S. acknowledge funding from the
Sherman Fairchild Foundation and by NSF Grants PHY-2011961, PHY-2011968,
and OAC-2209655 at Caltech.
\end{acknowledgments}

\appendix
\section{The two-grid method and transformations}
\label{app:two_grid_transformation}

In the Einstein frame, we adopt the 3+1 decomposition of the metric \cite{Baumgarte:2010ndz}
\begin{align}
ds^2=-\bar{\alpha}^2dt^2+\bar{\gamma}_{ij}(dx^i+\bar{\beta}^idt)(dx^j+\bar{\beta}^jdt),
\end{align}
where $\bar{\alpha}$, $\bar{\beta}^i$, $\bar{\gamma}_{ij}$ are the
lapse, shift, and 3-metric in the Einstein frame. They, their spatial
derivatives, and the extrinsic curvature $K_{ij}$ are transformed to the Jordan frame via:
\begin{equation}
\begin{aligned}
    &\alpha = \frac{1}{\sqrt{\phi}} \, \bar{\alpha}, ~~~
    \beta^i= \bar{\beta}^i, ~~~
    \gamma_{ij}= \frac{1}{\phi} \, \bar{\gamma}_{ij}, ~~
    \gamma^{ij} = \phi \, \bar{\gamma}^{ij}, \\
    &K_{ij} = \frac{1}{\sqrt{\phi}} \left( \bar{K}_{ij}
        + \frac{\bar{\gamma}_{ij}}{2} \, \frac{d\log{\phi}}{d\psi}
            \, \bar{n}^{k} \, \partial_k \psi
        \right), \\
    &\partial_k \alpha = \frac{1}{\sqrt{\phi}} \left( \partial_k \bar{\alpha}
        - \frac{\bar{\alpha} }{2} \, \frac{d\log{\phi}}{d\psi}\, \partial_k \psi
        \right) , \\
    &\partial_k \beta^i= \partial_k \bar{\beta}^i, \\
    &\partial_k \gamma^{ij} = \phi \left( \partial_k \bar{\gamma}^{ij}
        + \bar{\gamma}^{ij} \, \frac{d\log{\phi}}{d\psi} \, \partial_k \psi
        \right),
\end{aligned}
\end{equation}
where the future-directed unit timelike normal is given by
\begin{align}
\bar{n}^a=\bar{\alpha}^{-1} (\partial^a_t-\bar{\beta}^i\partial_i^a).
\end{align}
On the other hand, the transformation of the stress-energy tensor $\bar{T}^{ab}$ can be established from its definition
\begin{align}
    \bar{T}^{ab} = \frac{2}{\sqrt{-\bar{g}}} \frac{\delta S_M}{\delta
\bar{g}_{ab}}. \label{eq:Tbar_def}
\end{align}
After inserting
\begin{equation}
\begin{aligned}
    &\bar{g}_{ab}=\phi g_{ab},\\
    & \sqrt{-\bar{g}}=\phi^2\sqrt{-g},
\end{aligned}
\end{equation}
into Eq.~\eqref{eq:Tbar_def}, we obtain
\begin{align}
    \bar{T}^{ab} = \frac{2}{\sqrt{-\bar{g}}} \frac{\delta S_M}{\delta
\bar{g}_{ab}} = \frac{1}{\phi^3}\frac{2}{\sqrt{-g}} \frac{\delta S_M}{\delta
g_{ab}}= \frac{1}{\phi^3} T^{ab},
\end{align}
which leads to $\bar{T}_{ab}=T_{ab}/\phi$.

\section{Structure of neutron stars in ST gravity}
\label{app:stellar_structure}
Following Ref.~\cite{PhysRevLett.70.2220}, the Einstein-frame metric of an isolated, nonspinning NS can be written as
\begin{align}
    d\bar{s}^2=-e^{\nu(r)}dt^2+\frac{dr^2}{1-2\mu(r)/r}+r^2(d\theta^2+\sin^2\theta d\phi^2).
\end{align}
Then the equations of motion are given by
\begin{subequations}
  \label{eq:DEFsys}
  \begin{align}
    \mu' &= 4\pi  r^{2} A^{4} (\rho_0 h-P) + \frac{1}{2}r(r-2\mu) \varphi^{2},\label{eq:DEFsys-mu} \\
    \nu' &= 8\pi \frac{r^{2} A^{4}P}{r-2\mu} + r\varphi^{2} + \frac{2\mu}{r(r-2\mu)}, \label{eq:DEFsys-nu}\\
    \psi' &= \frac{1}{\sqrt{4\pi}}\varphi, \\
    \varphi' &= 4\pi \frac{r A^{4}}{r-2\mu}
    \left[
      (\alpha_0+\beta_0\sqrt{4\pi}\psi) (\rho_0 h-4P)\right. \notag \\
    &\left.+r\varphi(\rho_0 h-2P)
    \right] - \frac{2(r-\mu)}{r(r-2\mu)}\varphi, \\
    P' &= - \rho_0 h
    \left[
      \frac{1}{2}\nu'
      +(\alpha_0+\beta_0\sqrt{4\pi}\psi)\varphi
    \right],
  \end{align}
\end{subequations}
with $A=\phi^{-1/2}$. Note that $P$, $\rho_0$, and $h$ are in the
Jordan frame.  The system of coupled ordinary differential equations
can be solved as an initial value problem integrating out from
$r=\epsilon>0$.  The asymptotic expansion of the solution near the
stellar center $r\to 0$ is
\begin{equation}
  \begin{aligned}
      \mu(r) &\sim\frac{1}{3!}\mu_3r^3, \\
      \nu(r) &\sim\frac{1}{2!}\nu_2r^2, \\
      \varphi(r) &\sim\varphi_1 r, \\
      \psi(r) &\sim\psi_c+\frac{1}{2!}\frac{1}{\sqrt{4\pi}}\varphi_1 r^2, \\
      P(r) &\sim P_c+\frac{1}{2!}P_2r^2,
  \end{aligned}
\end{equation}
where
\begin{equation}
\begin{aligned}
    &\mu_3=8\pi A_c^4(\rho_ch_c-P_c), \\
    &\nu_2=8\pi A_c^4P_c+\frac{\mu_3}{3}, \\
    &\varphi_1=\frac{4\pi}{3}A_c^4(\alpha_0+\beta_0\sqrt{4\pi}\psi_c)(\rho_c h_c-4P_c),\\
    &P_2=-\rho_ch_c\left[\frac{1}{2}\nu_2+(\alpha_0+\beta_0\sqrt{4\pi}\psi_c)\varphi_1\right].
\end{aligned}
\end{equation}
We start the integration of Eqs.~\eqref{eq:DEFsys} at $\epsilon=
10^{-7}R^{\rm E}_{\rm ST}$
away from the stellar center, and terminate at the stellar
surface. From surface values, we obtain the scalar charge of the NS via \cite{PhysRevLett.70.2220}
\begin{align}
    \alpha_{\rm NS}=\left.\frac{2\varphi}{\nu^\prime}\right|_{\rm surf.}, \label{eq:scalar_charge_def}
\end{align}
and the Einstein-frame mass
\begin{align}
    m^{\rm E}_{\rm NS}=&\exp\left[-\frac{1}{\sqrt{1+\alpha_{\rm NS}^2}}{\rm arctanh}\left(\frac{\sqrt{1+\alpha_{\rm NS}^2}}{1+2/(r\nu^\prime)}\right)\right]\notag \\
    &\times\left.\frac{r^2\nu^\prime}{2}\left(1-\frac{2\mu}{r}\right)^{1/2}\right|_{\rm surf.}.
\end{align}
It is related to the Jordan-frame mass through \cite{Damour:1992we}
\begin{align}
    m^{\rm J}_{\rm NS}= m^{\rm E}_{\rm NS}(1+\alpha_0\alpha_{\rm NS}).
\end{align}

\begin{figure*}[tb]
        \subfloat[Scalar modes]{\includegraphics[width=0.9\textwidth,clip=true]{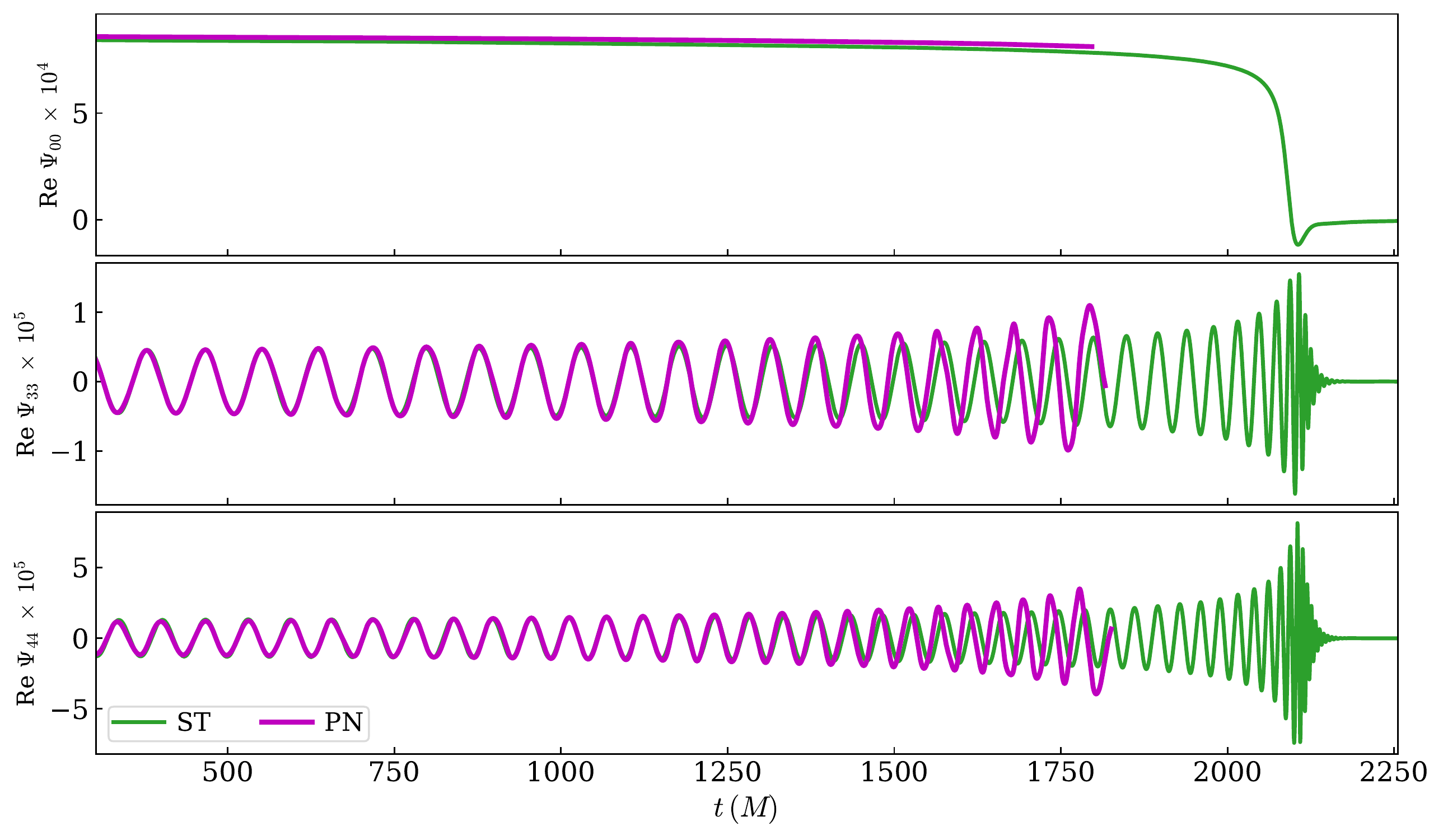}} \\
        \subfloat[Tensor modes]{        \includegraphics[width=0.9\textwidth,clip=true]{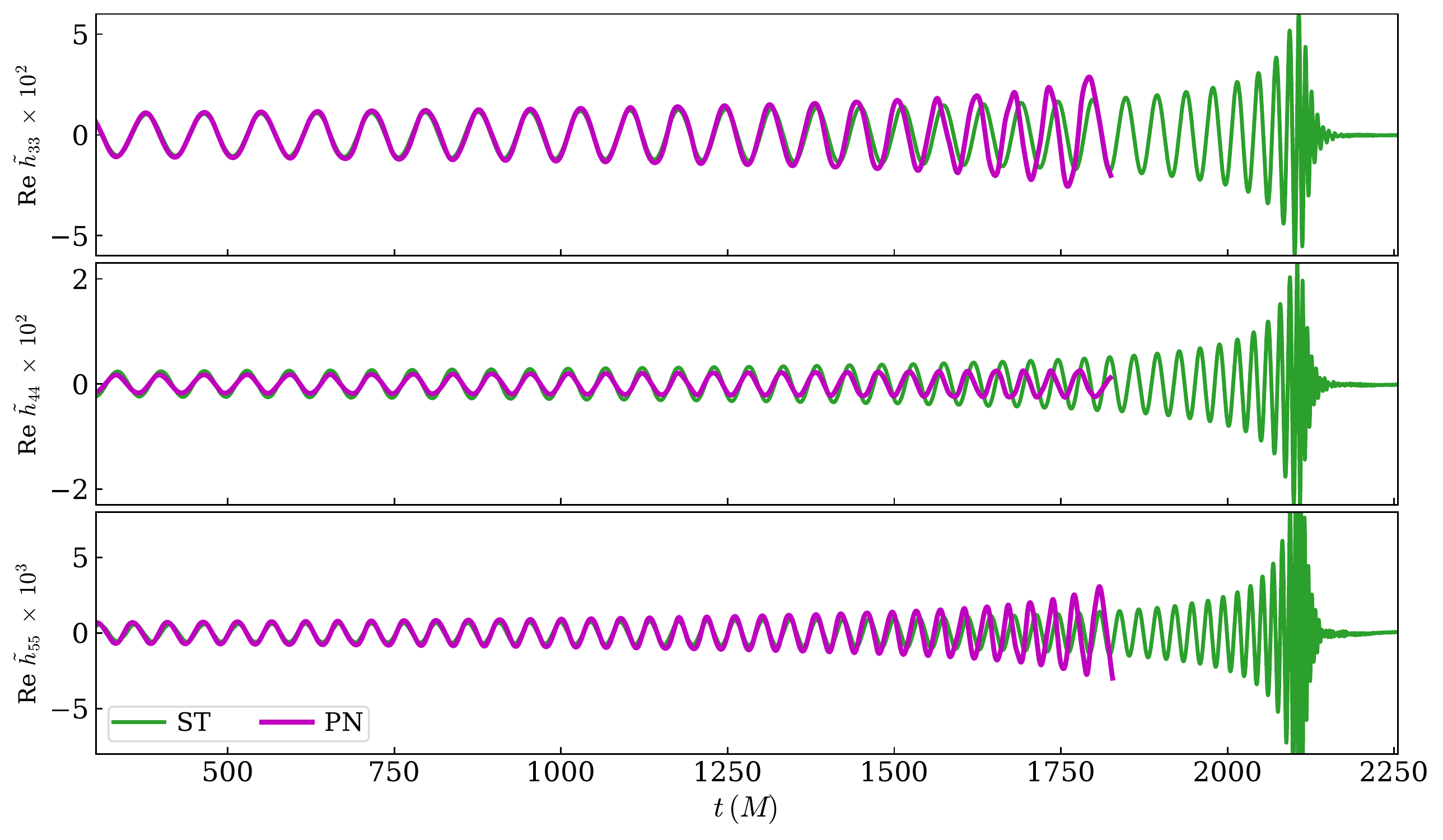}}
  \caption{Same as Fig.~\ref{fig:GR_PN}, some other scalar and tensor modes. Note that the modes $\Psi_{lm}$ are defined in Eq.~\eqref{eq:Psilm_psilm}.}
 \label{fig:higher_modes}
\end{figure*}

\begin{figure*}[tb]
  \includegraphics[width=\textwidth,clip=true]{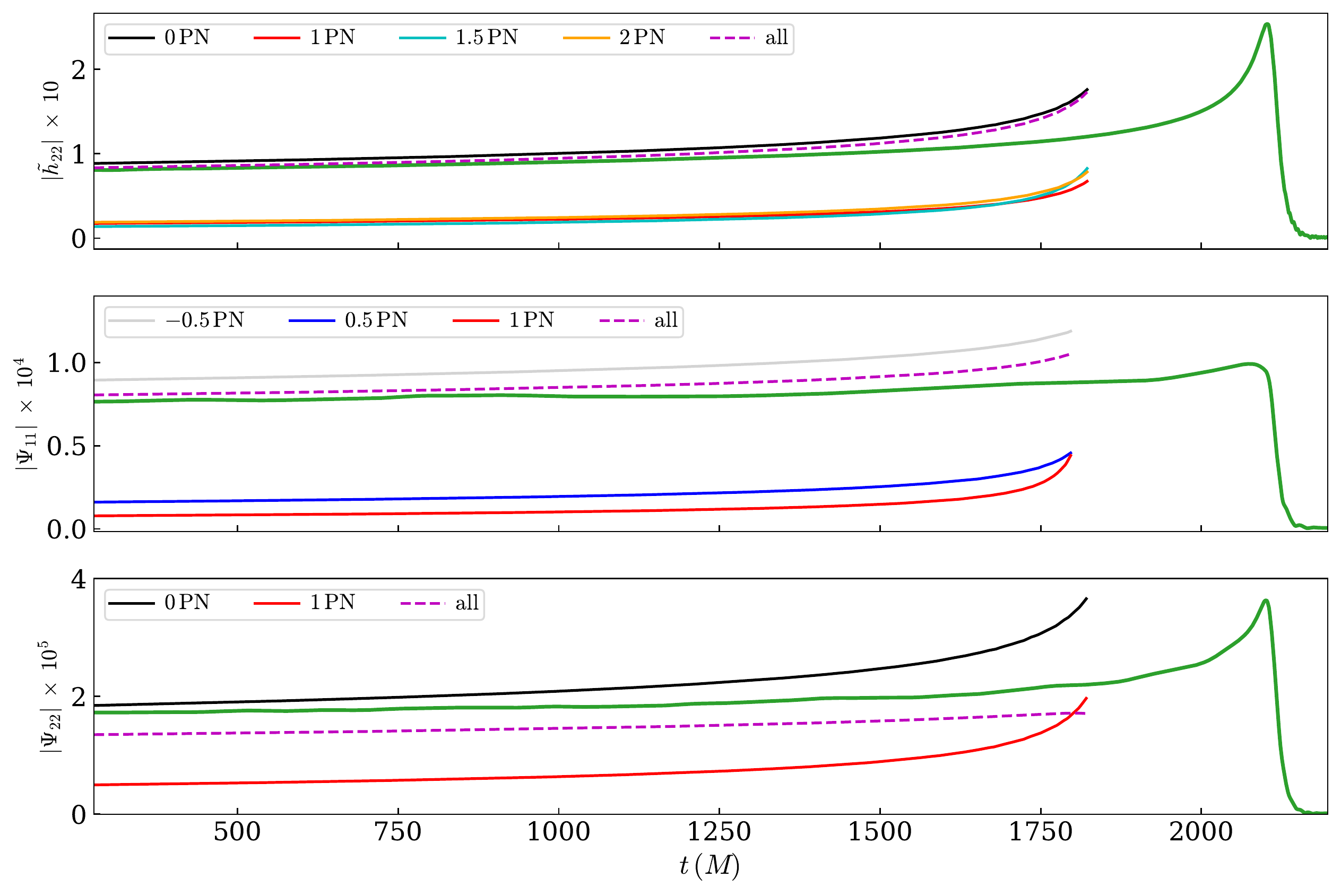}
  \caption{Contributions of individual PN orders to the amplitude of $\tilde{h}_{22}$ (top), $\Psi_{11}$ (middle), and $\Psi_{22}$ (bottom). It is important to note that a PN curve only includes contributions from the specific PN order, not lower PN orders. The magenta dashed curves refer to the ones that include all the PN terms.
  }
 \label{fig:PN_hierarchy_amp}
\end{figure*}

For a Newtonian star, Eqs.~\eqref{eq:DEFsys} reduce to
\begin{subequations}
\begin{align}
        \mu' &= 4\pi  r^{2} A^{4}(\psi_\infty) \rho_0, \label{eq:newton_mu}\\
        P'&=-\frac{\rho_0\mu}{r^2},
\end{align}
\end{subequations}
where the scalar field $\psi$ decouples from the matter and it becomes constant across the star. Here we denote its (background) value as $\psi_\infty$.
Next we can compute the baryonic mass $m^B$ and the Einstein-frame mass $m^{\rm E}_{\rm NS}$ of the NS:
\begin{subequations}
\begin{align}
    &m^B=A^3(\psi_\infty)\int 4\pi\rho_0 r^2dr, \\
    &m^{\rm E}_{\rm NS}=A^4(\psi_\infty)\int 4\pi\rho_0 r^2dr=m^BA(\psi_\infty). \label{eq:mb-mE}
\end{align}
\end{subequations}
As shown in Refs.~\cite{Damour:1992we,PhysRevLett.70.2220}, the scalar charge can be computed alternatively through
\begin{align}
    \alpha_{\rm NS}&=\frac{1}{\sqrt{4\pi}}\left(\frac{\partial \ln m^{\rm E}_{\rm NS}}{\partial \psi_\infty}\right)_{m^B}.
\end{align}
After plugging Eq.~\eqref{eq:mb-mE}, we obtain $\alpha_{\rm NS}=\alpha_0$ [see Eq.~\eqref{phi_psi}].

\section{Some other scalar and tensor modes}
\label{app:other_modes}
Figure \ref{fig:higher_modes} displays additional scalar and tensor modes of the ST simulation.

\section{Hierarchical contributions from PN terms}
\label{app:Hierarchical}
In Fig.~\ref{fig:GR_PN}, we compared the ST waveforms with the
existing PN predictions that include all the PN orders. Exploring the
hierarchical contributions of each PN term is also an interesting
aspect to investigate. Here we focus on the amplitude of
$\tilde{h}_{22}$ [Eq.~\eqref{hlm-GW}], $\Psi_{11}$ and $\Psi_{22}$
[Eq.~\eqref{scalar_mode_pn}], which are controlled by $\hat{H}_{lm}$ and $\hat{\Phi}_{lm}$ \cite{Sennett:2016klh}. Table~\ref{table:PN_order} outlines all the relevant PN orders of
$\tilde{h}_{22}$, $\Psi_{11}$, and $\Psi_{22}$. Our convention considers the leading Newtonian quadrupole
approximation in GR, namely $\mathcal{O}(1)$ in $\hat{H}_{lm}$, as 0PN. In contrast, the prefactor of
Eq.~\eqref{scalar_mode_pn} is 0.5PN $(x^{1/2})$ lower than that of
Eq.~\eqref{hlm-GW}, thus the term $\mathcal{O}(1)$ in $\hat{\Phi}_{lm}$ represents $-0.5$PN. 

We depict the size of each PN term as solid lines with different colors in Fig.~\ref{fig:PN_hierarchy_amp}. For reference, the dashed lines represent the ones
with all the PN contributions. The lowest PN order contributes the most, while higher PN corrections improve consistency. The amplitude of $\Psi_{22}$ is the least accurate. Higher PN
terms may be needed to improve the agreement with numerical simulations.

\begin{table}
    \centering
\renewcommand\arraystretch{1.4}
    \caption{Summary of all the PN orders in the amplitude of $\tilde{h}_{22}$, $\Psi_{11}$, and $\Psi_{22}$.}
\begin{tabular}{>{\centering\arraybackslash}p{1.7cm}@{\hspace{0cm}}p{4.3cm}@{\hspace{0cm}}p{2.7cm}}
\hline\hline
      Modes &
      \multicolumn{1}{l}{Available PN orders} &
      \multicolumn{1}{l}{References} \\ \hline
     \rowcolor{lightgray} $\tilde{h}_{22}$ & 0PN, 1PN, 1.5PN, 2PN & Eqs.~(67) of \cite{Sennett:2016klh} \\ \hline
    $\Psi_{11}$ &$-0.5$PN, $0.5$PN, 1PN & Eqs.~(6.10b) of \cite{Bernard:2022noq} \\ \hline
     \rowcolor{lightgray} $\Psi_{22}$ &0PN, 1PN & Eqs.~(6.10c) of \cite{Bernard:2022noq} 
\end{tabular}
     \label{table:PN_order}
\end{table}

\def\bibsection{\section*{References}}
\bibliography{References}

\end{document}